\documentclass[sigconf]{acmart}
\AtBeginDocument{%
  \providecommand\BibTeX{{%
    \normalfont B\kern-0.5em{\scshape i\kern-0.25em b}\kern-0.8em\TeX}}}

\usepackage{graphicx}
\usepackage{listings}
\usepackage{makecell}
\usepackage{multirow}
\usepackage{graphicx}
\usepackage{svg}
\usepackage{xspace}
\usepackage{dsfont}
\usepackage{lipsum}
\usepackage{subfigure}
\usepackage{centernot}
\usepackage{diagbox}
\usepackage{tabularx}

\newcolumntype{P}[1]{>{\centering\arraybackslash}p{#1}}

\newcommand{\tool}{{\texttt{Video2Action}}\xspace}

\definecolor{amber}{rgb}{1.0, 0.49, 0.0}

\copyrightyear{2023}
\acmYear{2023}
\setcopyright{acmlicensed}\acmConference[UIST '23]{The 36th Annual ACM Symposium on User Interface Software and Technology}{October 29-November 1, 2023}{San Francisco, CA, USA}
\acmBooktitle{The 36th Annual ACM Symposium on User Interface Software and Technology (UIST '23), October 29-November 1, 2023, San Francisco, CA, USA}
\acmPrice{15.00}
\acmDOI{10.1145/3586183.3606778}
\acmISBN{979-8-4007-0132-0/23/10}

\begin{document}

%%
%% The "title" command has an optional parameter,
%% allowing the author to define a "short title" to be used in page headers.
\title{Video2Action: Reducing Human Interactions in Action Annotation of App Tutorial Videos}

\author{Sidong Feng}
\affiliation{%
  \institution{Monash University}
  \city{Melbourne}
  \country{Australia}}
\email{sidong.feng@monash.edu}

\author{Chunyang Chen}
\affiliation{%
  \institution{Monash University}
  \city{Melbourne}
  \country{Australia}}
\email{chunyang.chen@monash.edu}
\authornote{Corresponding author}

\author{Zhenchang Xing}
\affiliation{%
  \institution{CSIRO’s Data61}
  \city{Canberra}
  \country{Australia}}
\email{zhenchang.xing@data61.csiro.au}

%%
%% By default, the full list of authors will be used in the page
%% headers. Often, this list is too long, and will overlap
%% other information printed in the page headers. This command allows
%% the author to define a more concise list
%% of authors' names for this purpose.
\renewcommand{\shortauthors}{Feng et al.}

%%
%% The abstract is a short summary of the work to be presented in the
%% article.
\begin{abstract}
  Tutorial videos of mobile apps have become a popular and compelling way for users to learn unfamiliar app features.
To make the video accessible to the users, video creators always need to annotate the actions in the video, including what actions are performed and where to tap.
However, this process can be time-consuming and labor-intensive.
In this paper, we introduce a lightweight approach \tool, to automatically generate the action scenes and predict the action locations from the video by using image-processing and deep-learning methods.
The automated experiments demonstrate the good performance of \tool in acquiring actions from the videos, and a user study shows the usefulness of our generated action cues in assisting video creators with action annotation.
\end{abstract}

%%
%% The code below is generated by the tool at http://dl.acm.org/ccs.cfm.
%% Please copy and paste the code instead of the example below.
%%

%%
%% Keywords. The author(s) should pick words that accurately describe
%% the work being presented. Separate the keywords with commas.

%%
%% This command processes the author and affiliation and title
%% information and builds the first part of the formatted document.
\maketitle

\section{Introduction}
Mobile apps now have become the most popular way of accessing the Internet as well as performing daily tasks, e.g., reading, shopping, banking, and chatting~\cite{chen2019gallery,feng2022gallery}.
It is not always clear to a user how to access specific functionalities, such as posting the event on the social app or enabling tethering on the collaboration app.
Consequently, mobile users usually look for tutorials on how to perform a specific task on the web and social media, such as Youtube~\cite{web:youtube}, wikiHow~\cite{web:wikihow}, etc.

It is common for app tutorials to be presented in the form of text-based documents~\cite{zhong2021helpviz}. However, an increasing number of tutorials are now being created as screen recordings~\cite{feng2022gifdroid,feng2022gifdroid1}.
On the one hand, compared to writing clear and concise documentation, video-based tutorial significantly lowers the bar for tutorial creators.
It is easy to record the screen as there are many tools available, some of which are even embedded in the operating system by default like iOS~\cite{web:iosrecord} and Android~\cite{web:androidrecord}.
On the other hand, the video-based tutorial is more immersive and engaging for users to learn the unfamiliar apps~\cite{craig2013video} and it can include more detail and context, such as configurations, and parameters, to help users replicate the app functionalities on their own device.

Despite their merits, video-based tutorials still pose a challenging context for users due to the lack of accessibility.
First, the video may play too fast to watch, and the actions performed in the video sometimes are too inconspicuous to be realized~\cite{web:why_video,kafer2017best,digmayer2012help,feng2023read}.
For example, users may need to carefully observe fast UI movements during scrolling, recognize tap locations by small and unobtrusive UI transition animations, and memorize previous UIs for backward action identification.
This can be time-consuming and cognitively demanding, especially for new or older users if they have little experience with the app or are not familiar with the specific actions being performed.
Second, while the auditory information in the video (i.e., narration) may help users understand the actions, it is not accessible to non-native-language users or hearing-impaired users~\cite{gernsbacher2015video,zhang2021using}.
These challenges make it difficult for users to understand and follow along with the tutorials, and users may not always be able to trial and error to figure out what's going on in a tutorial.

\begin{figure}[htp!]
	\centering
	\includegraphics[width = 0.82\linewidth]{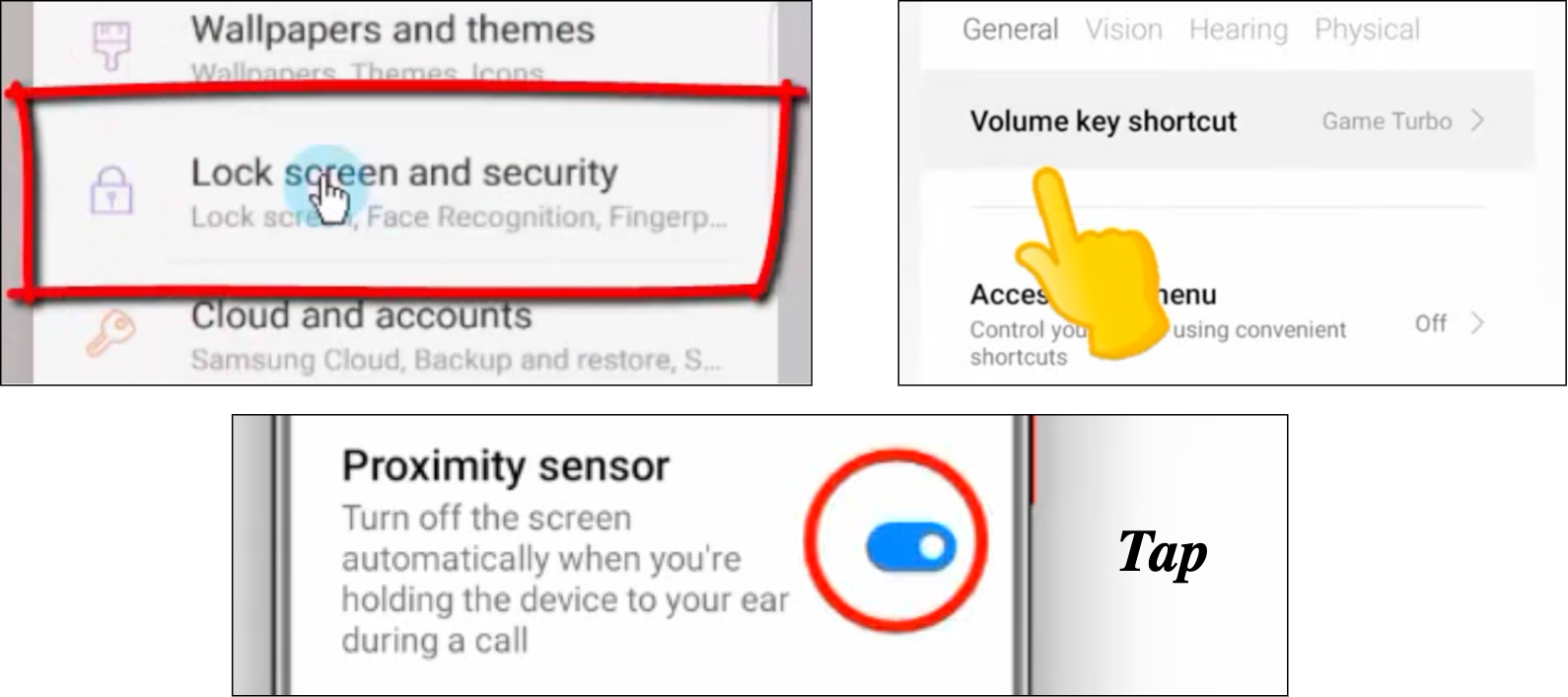}
	\caption{Examples of action annotations in tutorial videos.
	}
	\label{fig:touchExample}
\end{figure}

To address this, video creators often utilize the markers (such as bounding boxes, action illustrations, etc.) to explicitly highlight the action cues in the video (see Figure~\ref{fig:touchExample}), guiding the user’s attention to essential actions and UI elements~\cite{van2022eleven,van2013eight}.
This can reduce the extraneous cognitive load associated with locating relevant information and can free up mental resources allocated to understanding the content.
However, manually annotating the app tutorial videos can be time-consuming and labor-intensive for video creators, including watching the video frame-by-frame, extracting the action clips, recalling the specific action locations, and annotating the actions.
There are many studies that attempt to facilitate the annotation of natural videos~\cite{wang2021soloist,zhang2020rcea,piazentin2019historytracker,deng2021eventanchor,shang2019annotating}, but rarely related to annotating the actions in mobile app videos.
Some researchers model mobile app UIs and UI interactions based on a single static UI~\cite{liu2018learning,swearngin2019modeling,chen2019gallery,feng2022auto}, however, those approaches do not apply to model semantic interactions on video artifacts (sequences of UIs).
To retrieve the action execution information, software instrumentation~\cite{nurmuradov2017caret, qin2016mobiplay} is widely used, i.e., adding extra code to an app for monitoring UI interactions. 
However, instrumentation requires sophisticated accessibility or UI automation APIs~\cite{grechanik2009creating,hao2014puma} and continuous updates along with the app and different operating systems~\cite{li2018cid,wang2013detecting}.
In addition, the intrusive techniques cannot reliably and accurately acquire information from apps~\cite{li2022learning,li2020mapping}, e.g., misaligned runtime view hierarchy. 
Some studies work on extracting the actions from the app usage videos based on extra recording apparatus, such as developer-mode touch indicators~\cite{bernal2020translating}, third-party screen recorders~\cite{web:az}, or external cameras ~\cite{qian2020roscript}.
These works add extra work for video creators, but not all non-developer or non-tester creators have such domain knowledge and are willing to use it according to our empirical study in Section~\ref{sec:background}.

In this paper, we present \tool, a lightweight non-intrusive approach that only requires an app tutorial video as the input and automatically acquires the actions from the video, enabling human-AI collaboration to reduce the burdens of video creators in action annotation.
Our approach consists of two main phases: 1) Action Scene Generation and 2) Action Location Prediction.
First, we propose a heuristic image-processing method to segment the app video into action scenes.
Given the action scenes, we then develop a novel deep-learning method to infer the action locations.
Based on the actions acquired by our approach, we further implement a proof-of-concept user interface to offer an opportunity for video creators to navigate to specific frames of actions in the video, identify action locations, and effectively create annotations.

We evaluate our approach \tool based on a large-scale crowdsourced Rico~\cite{deka2017rico}.
Results show that our approach achieves the best performance (81.6\% Video F1-score and 86.4\% Levenshtein score) in action scene generation from the videos compared with six commonly-used baselines. 
Our approach also achieves on average 50.1\% and 81.9\% accuracy in inferring top-1 and top-5 action locations, which significantly outperform three state-of-the-art baselines and three ablation studies. 
We further carry out a user study to evaluate the usefulness of \tool in assisting action annotation of app tutorial videos in the real-world environment.
Results show that participants save 85\% of time annotating the actions with the help of the actions generated by our approach, in comparison to the annotation from scratch.
The feedback from the participants also confirms the usefulness and helpfulness of the \tool in the social media community.
Finally, we discuss the generality of our approach and show two potential applications that could benefit from our approach to interact or collaborate with, including bug recording replay and video captioning.

The contributions of this paper are as follows:

\begin{itemize}
  \item We present a lightweight non-intrusive approach \tool for automatically acquiring actions from the app tutorial video to reduce human interaction burdens in action annotation.
  \item We conduct an empirical study to investigate the action annotation problems of the app tutorial videos and understand the characteristic of actions.
  \item A comprehensive evaluation including automated experiments and a user study to demonstrate the performance and usefulness of \tool.
\end{itemize}

\section{Related Work}

\subsection{Annotating UI-based Videos}
The advance of machine learning has provided new opportunities to reduce the cognitive and interaction burdens of users in video annotation,
such as an adaptive video playback tool to assist the quick review of long video clips~\cite{al2020fast,wang2021soloist}, a mobile application to support real-time, precise emotion annotation~\cite{zhang2020rcea}, an interaction pipeline for the annotation of objects and their relations~\cite{shang2019annotating}, and a novel method to acquire tracking data for sports videos~\cite{piazentin2019historytracker,deng2021eventanchor}.
These prior studies focus on the videos of natural scenes or virtual scenes, and cannot easily transfer to our domain of digital scenes, UI screen-casting videos.

There are few researchers that work on desktop-based screen-casting in assisting software development~\cite{bao2017extracting, banovic2012waken, chen2022extracting}.
In contrast, we focus on the recording of more compact and denser screen, mobile-based videos.
Most of the work for mobile videos is to facilitate automated app testing by bug record-and-replay, which aims to capture the screens that triggered the bug and play it back on the device.
For example, Nurmuradov et al.~\cite{nurmuradov2017caret} developed a program analysis tool to dump the action data during the recording process and then replay the actions on an Android emulator.
The underlying technique of these works is software instrumentation by adding extra code to an app for monitoring action behavior.
However, it relies on sophisticated accessibility or UI automation APIs (i.e., Accessibility, replaykit)~\cite{grechanik2009creating,hao2014puma} and continuous updates along with the app and different operating systems~\cite{li2018cid,wang2013detecting}.
In many cases, the intrusive techniques cannot reliably and accurately acquire information from the apps~\cite{li2022learning,li2020mapping}, i.e., misaligned runtime view hierarchy.
Bernal et al.~\cite{bernal2020translating} introduced a lightweight record-and-replay tool V2S, but it required testers to access the Android developer setting to enable the touch indicator for action identification.
A similar work is RoScript~\cite{qian2020roscript} which required testers to use an external camera to record the screen and finger movement.
These works add extra work for video creators to record the screen, but not all non-developer or non-tester creators have such domain knowledge and are willing to spend that much effort according to our empirical study in Section~\ref{sec:background}.

In this study, we propose a purely image-based approach to acquire the actions from the app tutorial videos, without any requirement of heavy testing framework installation, developer configuration setup, or extensive app instrument.
In detail, we first leverage the image-processing method to segment the video into action scenes and then adopt deep-learning models to infer the action locations.
With the rich action information acquired by our approach, we support efficient video content exploration, thus significantly reducing the burdens of video creators in action annotation.

\subsection{Modeling UI Interactions}
\tool is related to prior research on computationally modeling app UIs and UI interactions~\cite{feng2023prompting,liu2018learning,feng2023efficiency,chen2020lost,xie2020uied}.
For example, Swearngin et al.~\cite{swearngin2019modeling} proposed a machine learning method TapShoe, that leveraged the tappability signifiers of UI (e.g., type, size, text) to model whether the UI elements are tappable.
Unlike these works, which model UI interactions using information from a single static UI image, we model interactions based on UI response, i.e., recognize the actions triggered from one UI to the next, allowing for more advanced semantic interactions, such as scrolling the UI or returning to the previous UI, etc.

Lee et al.~\cite{lee2018click} developed a method to predict the UI element that the user is likely to tap on the current screen based on the previous screens.
The underlying technique was a sequence model LSTM that treats actions as a sequence of tokens derived from the UI elements.
A similar work is Humanoid~\cite{li2019humanoid} which modeled UI interactions as automated UI testing.
With the emergence of the Transformer~\cite{vaswani2017attention}, Chen et al.~\cite{chen2019behavior} leveraged the element information from the UI hierarchy to train a Transformer to recommend the next tap location in the shopping app.
Given the multi-modal information such as the user’s action history and time of the day, Zhou et al.~\cite{zhou2021large} further improved the performance of tap location prediction.
In contrast, our work focuses on the screenshots from app tutorial videos, which are just UI images without additional information.
He et al.~\cite{he2021actionbert} introduced an image-processing method ActionBert to predict the tap location between UI images.
The pipeline of ActionBert was to first detect the elements in the UI, then extract their information, and finally predict the tap location.
However, this step-by-step method can lead to a ``garbage in and garbage out'' problem, i.e., imprecise UI element detection will result in incorrect tap location prediction.
To that end, we propose an end-to-end differentiable model to detect the potential region of interest (i.e., tappable elements) in the UI, and extend them to infer the specific tap location to trigger the next UI.
Considering the human knowledge of UI and UI interaction, we further develop a tailored data-augmentation method to enhance the robustness of our model.
The results in Section~\ref{sec:eval_location} demonstrate our model with deep human knowledge can achieve better performance in modeling UI interaction.

\section{Empirical Study of App Tutorial Videos and Related Problems}
\label{sec:background}
In this section, we carry out a small empirical study to gain insight into the app tutorial videos and find implicit characteristics of user interaction behaviors for motivating the required tool support.
We select Youtube~\cite{web:youtube} as our study subject as it is the most widely-used social media sharing platform.
We randomly collect 500 videos as our experimental dataset,
showcasing a variety of tutorial topics, including instructional content and app demo walkthroughs, recorded through screencasting. Among these, 69\% feature narration, while 44\% are in English.

To gain an understanding of app tutorial videos, we recruit three labelers from online posting to label the actions from our experimental dataset.
All labelers have more than two years of labeling experience and have labeled at least one UI/UX-related dataset (i.e., UI element bounding box, UI animation linting).
We first give them an introduction to our study and also a real example to try.
Then, we assign the experimental set of app tutorial videos to them to label the actions from a comprehensive list of user interactions~\cite{web:gesture} independently without any discussion.
After the initial labeling, the labelers meet and correct the subtle discrepancies.
In total, we obtain 9,764 actions from 500 experimental app tutorial videos, on average 19.5 actions per 3.2 minutes video. 
Based on the action labels, the following research questions are emerged.

\subsection{Are the actions in the app tutorial videos clearly annotated?}
We observe that more than 62\% of videos are without action cues.
To confirm such a phenomenon and understand the limitations of action annotation, we further conduct informal interviews with three professional app tutorial creators, including two creators (C1, C2) from the Alibaba app development team, and one creator (C3) with more than 10k followers on Youtube.
All of them mention that they do annotate the actions for some important tutorials (i.e., app releases, key feature introductions, etc.) and acknowledge that it can engage users and gain more attention from the community.
However, they may not annotate every video due to the following practical reasons.
First, annotating actions in the video is a time-consuming and tedious process.
As C2 says: \textit{``To create a great tutorial video with action annotations, I need to first split the video into action clips based on the timing of each action. Since the actions in the video may play too fast, I always need to pause and replay the video multiple times. And sometimes I need to watch the video frame-by-frame to split the action clips precisely. After getting the clips, I need to further recall the action attributes such as the tapping location, the scrolling offset, etc., and finally, use the video editing tools to annotate the actions.''}
C1 also confirms the challenges of adding action annotations in the industry due to budget constraints and market pressures. 

Second, developers and testers have developed built-in touch indicators~\cite{bernal2020translating} or third-party screen recorders~\cite{web:az} to annotate the actions performed on the screen for automated app testing.
However, the creators may not have the domain knowledge to set it up, as C3 says: 
\textit{``I tried following the developers' instructions to enable the default touch indicator on the device, but I found it too difficult, requiring opening developer settings, rebooting the device, etc.''}
In addition, C3 explains the inadequacy of such touch indicators for users replaying the videos and emphasizes the necessity of manual action annotation: 
\textit{``Annotation is meant to guide the user's attention to the key elements or locations on the UI.
However, existing touch indicators, such as built-in indicators or third-party cursors, are too small (less than 1\% of the UI), inconspicuous (low contrast with the UI), and unclear (they don't show action semantics).
As a result, they are not very helpful for users to learn and follow.
To help users perceive the key points of actions more easily, I will use high-contrast colors and well-sized annotations (such as arrows, bounding boxes, and action illustrations) to explicitly highlight the actions as shown in Figure~\ref{fig:touchExample}.''}

\subsection{What are the actions in the app tutorial video?}
\label{sec:background1}
Although the set of actions is fairly large~\cite{wobbrock2009user}, there is often only a limited set of actions that are appropriate for a given app and device.
To provide opportunities for all users to learn and replicate, the video creators often make the actions in the tutorial semantically clear and simple~\cite{de2009towards}.
For instance, the swipe action that moves from left to right to return to the previous UI, is not supported on older devices, and one simple alternative is to tap the system's backward button.
Therefore, we investigate the actions in the app tutorial videos to gain insight into common user interaction behaviors.
Across all the labeled actions (9,764) in the experimental dataset, we find three most commonly-used actions:

\begin{itemize}
  \item \textbf{\textsc{TAP} (80.4\%)} Allows users to interact with elements and access additional functionality. It usually transits to a very different UI.
  \item \textbf{\textsc{SCROLL} (10.7\%)} Allows users to slide screens vertically or horizontally to move continuously through content. 
  \item \textbf{\textsc{BACKWARD} (7.5\%)} A semantic action of tap returns to the previous screen. It is often used to return to the app's landing page to demonstrate the next app functionality. It can be done by tapping the backward button in the system navigation bar at the bottom of the screen.
  \item \textbf{Others. (1.4\%)} There are also some other actions such as pinch, flinch, etc. However, they rarely appear in our experimental dataset ($< 2\%$).
\end{itemize}

\begin{figure}
	\centering
	\includegraphics[width = 0.99\linewidth]{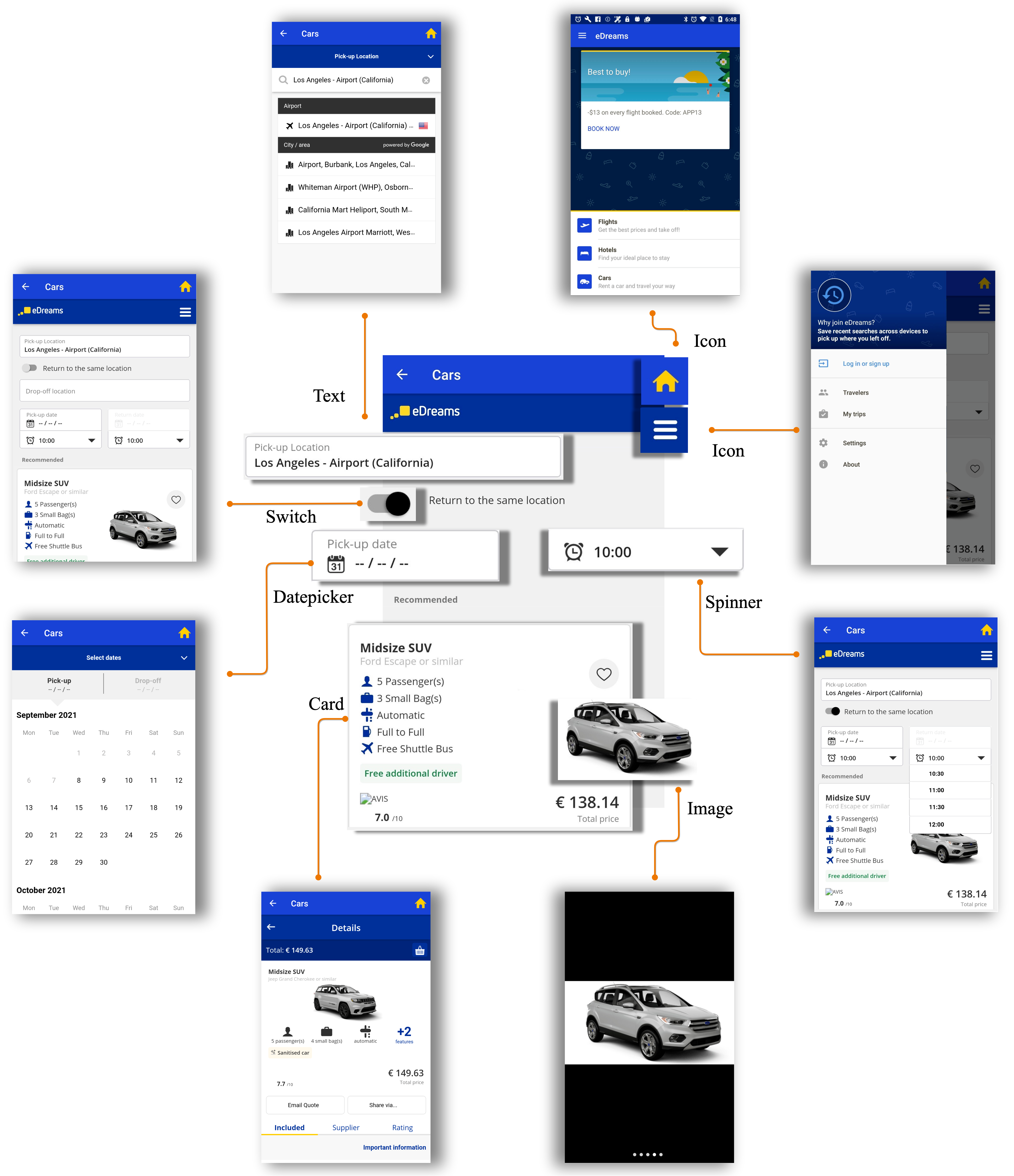}
	\caption{An illustration example of tapping area and corresponding response in the UI.}
	\label{fig:tapcategory}
\end{figure}

\renewcommand{\arraystretch}{1.05}
\begin{table*}
    \small
	\centering
	\caption{The categorization of tapping actions.}
	\label{tab:transition_category}
	\begin{tabularx}{\textwidth}{l|c|X} 
		\hline
		\normalsize{\bf{Tapping Category}} & \normalsize{\bf{Specific Transition}} & \normalsize{\bf{Description}} \\
		\hline
		\hline
		\multirow{5}{*}{TAPPING AREA} & Text & The transition is triggered by the text content of an element, such as text, text field. \\
		 \cline{2-3}
		 & Image & The transition is triggered by the image view of an element, such as image, icon. \\
		 \cline{2-3}
		 & Button & The transition is triggered by the essential interactive elements, including button, toggle button, radio button, multi-tab button, spinner, switch, and checkbox. \\
		\cline{2-3}
		 & Others & Some infrequently seen elements can also trigger UI transitions, such as rating bar, seek bar, etc.\\
		\hline
		\multirow{8}{*}{TAPPING RESPONSE} & New Page & It transits to a new UI. They may have some similar content, but they are visually different.  \\
		 \cline{2-3}
		 & Pop Up & It pops up a modal or dialog that appears on top of the previous UI with the background dimmed, such as tapping on the menu icon. \\
		 \cline{2-3}
		 & Dropdown Menus & Different from the pop-up response, it reveals a list of options or commands and keeps the context of the information being requested visible, such as tapping on the spinner button.\\
		\cline{2-3}
		 & Selection Control & It responds to the users with visual feedback when they control certain options, settings, or states in the selection buttons, such as radio button, switch, and checkbox. \\
		\cline{2-3}
		 & Others & There are also some potential responses for UI transitions, such as text input, video play, etc. We categorize them as Others as they rarely appear in our empirical dataset ($< 1\%$). \\
		 \hline
	\end{tabularx}
\end{table*}

\subsection{What are the potential patterns in \textsc{TAP} actions?}
\label{sec:pattern2}
As the dominant action, tapping involves more diverse elements and responses than other actions as shown in Figure~\ref{fig:tapcategory}. 
To further understand \textsc{TAP} actions, we ask three labelers to code the categories of tapping patterns using the existing UI/UX design knowledge documented in books and websites such as The Design of Everyday Things~\cite{norman2013design} and Mobile Design Pattern Gallery~\cite{neil2014mobile}.

According to the background of UI and the transitions that triggered, we define the key characteristics of tapping actions into two main categories as shown in Table~\ref{tab:transition_category}.
First, we identify the TAPPING AREA to describe the tapping location that triggers the UI transition.
Second, we define the TAPPING RESPONSE to describe the rendering effect after tapping.
Each of these main categories has a subset of specific categories, which jointly describe a tapping interaction. 
For example, as shown in the upper right of Figure~\ref{fig:tapcategory}, when the user taps the menu \textbf{icon}, it will transit to a UI with a \textbf{pop-up} menu list view.
Another example shown in the upper left of Figure~\ref{fig:tapcategory} illustrates an interaction by tapping the \textbf{text} view to a \textbf{new page} UI with different content and layout.

\vspace{12pt}
\noindent\fbox{
    \parbox{0.98\linewidth}{
        \textbf{Summary}: 
        By analyzing 500 app tutorial videos from Youtube, 62\% of them are without action annotation. 
        Despite the set of actions is fairly large, there are three most commonly-used actions in the tutorial videos, i.e., \textsc{TAP}, \textsc{SCROLL}, \textsc{BACKWARD}.
        As the most common action (85.4\%), \textsc{TAP} action involves diverse area and corresponding response, resulting in the difficulty in identifying the tap location in the screen even by a human.
    }
}

\section{\tool Approach}
\begin{figure*}
	\centering
	\includegraphics[width = 0.86\textwidth]{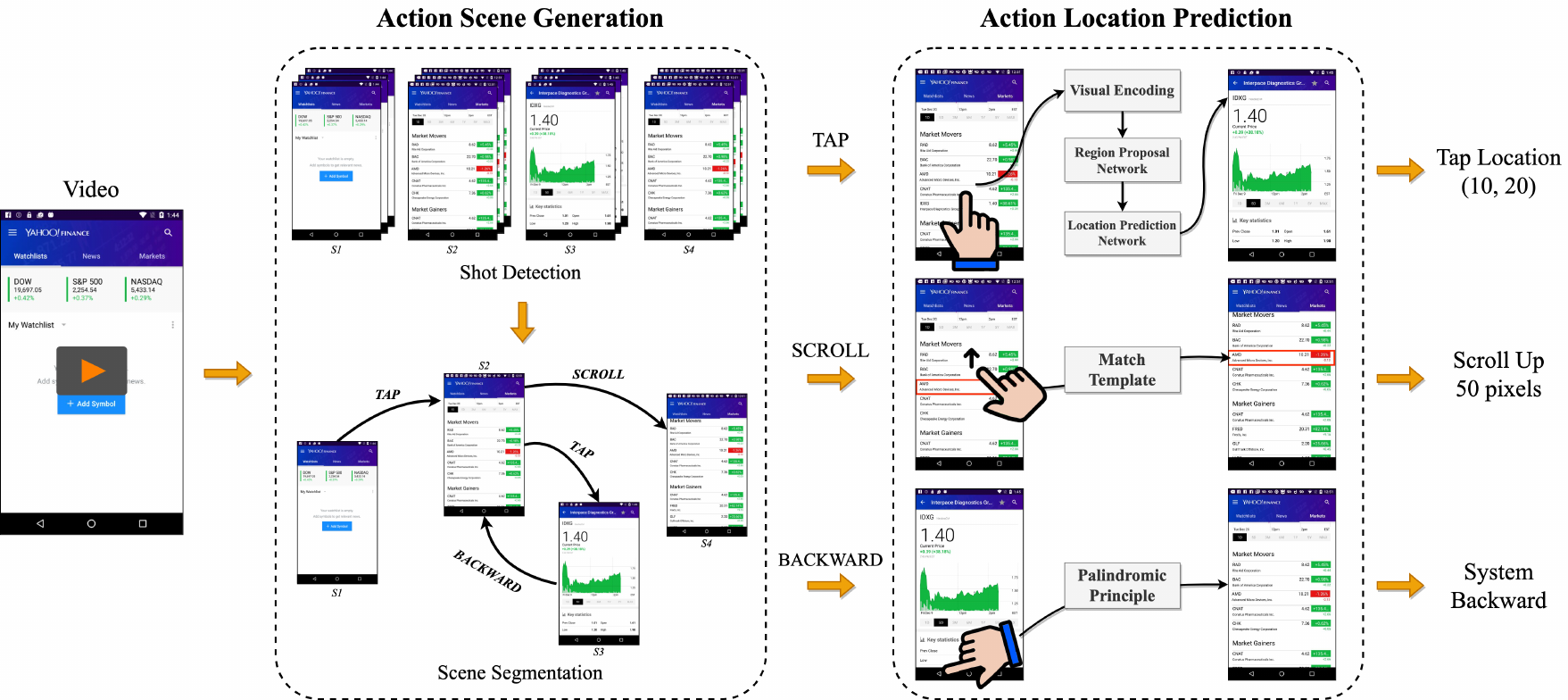}
	\caption{Overview of \tool to acquire actions from the video, consists of two main phases. \textbf{Action Scene Generation} phase takes a video as input and segments it into a scene transition graph with UI actions (e.g., Tap, Scroll, Backward). For each action, \textbf{Action Location Prediction} phase infers specific locations by adopting image-processing and deep-learning methods.}
	\label{fig:approach}
\end{figure*}

The findings in Section~\ref{sec:background} confirm the necessity and difficulty of annotating actions in the app tutorial videos and motivate our approach development for automatic action acquisition to significantly reduce the cognitive and interaction burdens of video creators in action annotation.
The overview of \tool is shown in Figure~\ref{fig:approach}, consists of two main phases: \textbf{Action Scene Generation} and \textbf{Action Location Prediction}.

For \textbf{Action Scene Generation}, since people perceive a sequence of graphics changes as a motion, consecutive images are perceptually dissimilar if people recognize any motions (i.e., UI transitions) from the image frames~\cite{tversky2002animation}.
In the human perception (a.k.a human vision) system, a majority of visual information is conveyed by patterns of contrasts from its brightness changes~\cite{zeki1993vision}.
Inspired by the biological vision, we propose a heuristic image-processing method based on brightness computation to segment action scenes from the video.
That is, we first compute the luminance similarity between consecutive frames and cut the video into shots.
Given the shots and consecutive frame similarity sequence, we then classify the action types (i.e., \textsc{TAP}, \textsc{SCROLL}, \textsc{BACKWARD}) and semantically correlate the shots into scenes.

For \textbf{Action Location Prediction}, we aim to infer the action locations between scenes.
For \textsc{SCROLL} action, we adopt the template matching~\cite{brunelli2009template} method to calculate the moving distances; for \textsc{BACKWARD} action, we utilize the built-in system backward button. 
Since these methods are well-known and well-implemented, we omit the details for brevity in this paper.
For \textsc{TAP} action, considering the diversity of tapping area and response observed in Section~\ref{sec:pattern2}, it would require significant effort to manually build a complete set of rules to detect action positions in all different situations. 
Therefore, we propose a novel deep-learning model to automatically learn the tapping area from the UI and predict the tapping coordinates.
To improve the robustness and performance of the model, we further apply a tailored data augmentation method and a post-processing technique.

\subsection{Action Scene Generation}
\label{sec:scene_generation}

\subsubsection{Shot Detection}
Different from natural scene videos, UI videos have clear shot boundaries of user actions, i.e. the start and end frames of a fully rendered UI.
To detect the shots, we leverage the image-processing techniques to build a perceptual similarity score for consecutive frame comparisons based on luminance difference Y-Diff in YUV color space.
Consider a video $\big\{ f_{0}, f_{1}, .., f_{N-1}, f_{N} \big\}$ , where $f_{N}$ is the current frame and $f_{N-1}$ is the previous frame.
To calculate the Y-Diff of the current frame $f_{N}$ with the previous $f_{N-1}$, we first obtain the luminance mask $Y_{N-1}, Y_{N}$ by splitting the YUV color space converted by the RGB color space.
Then, we apply the perceptual comparison metric, SSIM (Structural Similarity Index)~\cite{wang2004image}, to produce a per-pixel similarity value related to the local difference in the average value, the variance, and the correlation of luminances.
An SSIM score is a number between 0 and 1, and a higher value indicates a strong level of similarity.

\begin{figure}
	\centering
	\includegraphics[width = 0.93\linewidth]{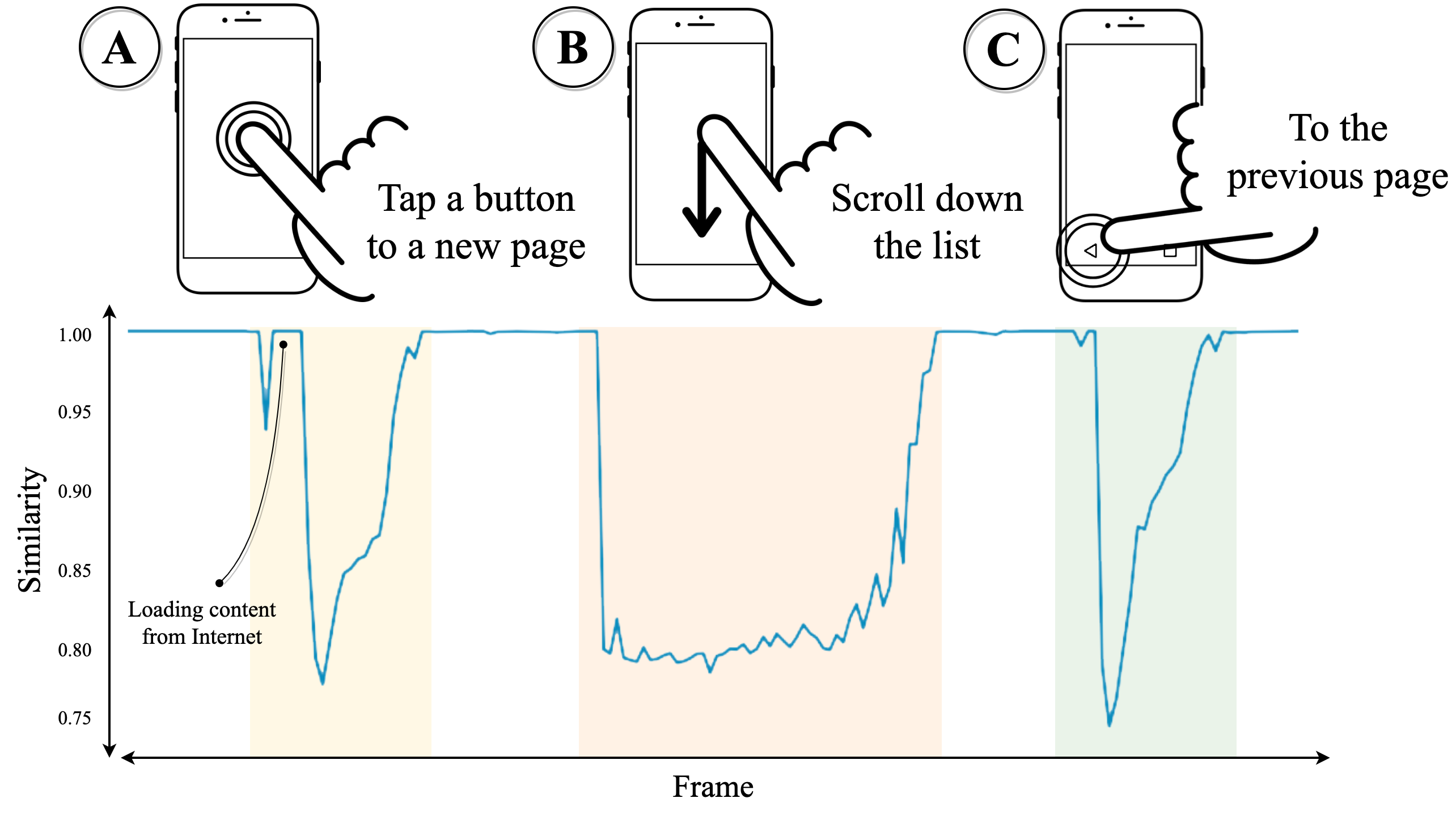}
	\caption{An illustration of Y-Diff similarity scores of consecutive frames in the UI video.
	}
	\label{fig:simExample}
\end{figure}

Figure~\ref{fig:simExample} shows a consecutive frame similarity sequence of a UI video.
A shot is selected to be the fully rendered UI, that is the steady state where the consecutive frames are similar for a relatively long duration.
The reason why we choose long duration is because of the occurrence of short steady duration in Figure~\ref{fig:simExample}A.
While the UI layout of UI rendering is fast, resource loading may take time. 
For example, rendering images from the web depends on device bandwidth, image loading efficiency, etc.
Based on a small pilot study, we set a duration of 1 second as a relatively long duration.

\subsubsection{Scene Segmentation}
Videos such as movies, documentaries, and TV-series, follow some production rules~\cite{chasanis2008scene} to proceed with shots to generate semantic correlated scenes.
To generate these rules in UI videos, we look into the similarity scores of consecutive frames and their corresponding shots as shown in Figure~\ref{fig:simExample}.
As we notice, the semantics of scenes strongly match the UI transition patterns observed in Section~\ref{sec:background1}.
Therefore, we develop a heuristic approach to identify the semantics of scenes following the matching patterns:

(1) \textsc{TAP}: 
usually instantly transits UI to a very different UI as discussed in Section~\ref{sec:background1}, revealing a drastically low similarity score during the transition, such as Figure~\ref{fig:simExample}A.

(2) \textsc{SCROLL}:
implicates a continuous transition from one UI to another, consequently, the similarity score starts with a drastically drop and then continues to increase slightly over a period of time, such as Figure~\ref{fig:simExample}B.

(3) \textsc{BACKWARD}:
depicts a semantic transition from the current UI to the previous UI as shown in Figure~\ref{fig:simExample}C.
However, the similarity score cannot reliably detect \textsc{BACKWARD} actions, as it may coincide with the \textsc{TAP} actions.
According to the \textsc{BACKWARD} actions are palindromic, e.g., UI-1 $\xrightarrow[]{\text{Tap}}$ UI-2 $\xrightarrow[]{\text{Tap}}$ UI-1, we develop a stack that follows the LIFO principle (last in, first out)~\cite{cormen2009introduction} to check whether the palindromic UI shots are identical.

\begin{figure*}
	\centering
	\includegraphics[width = 0.78\linewidth]{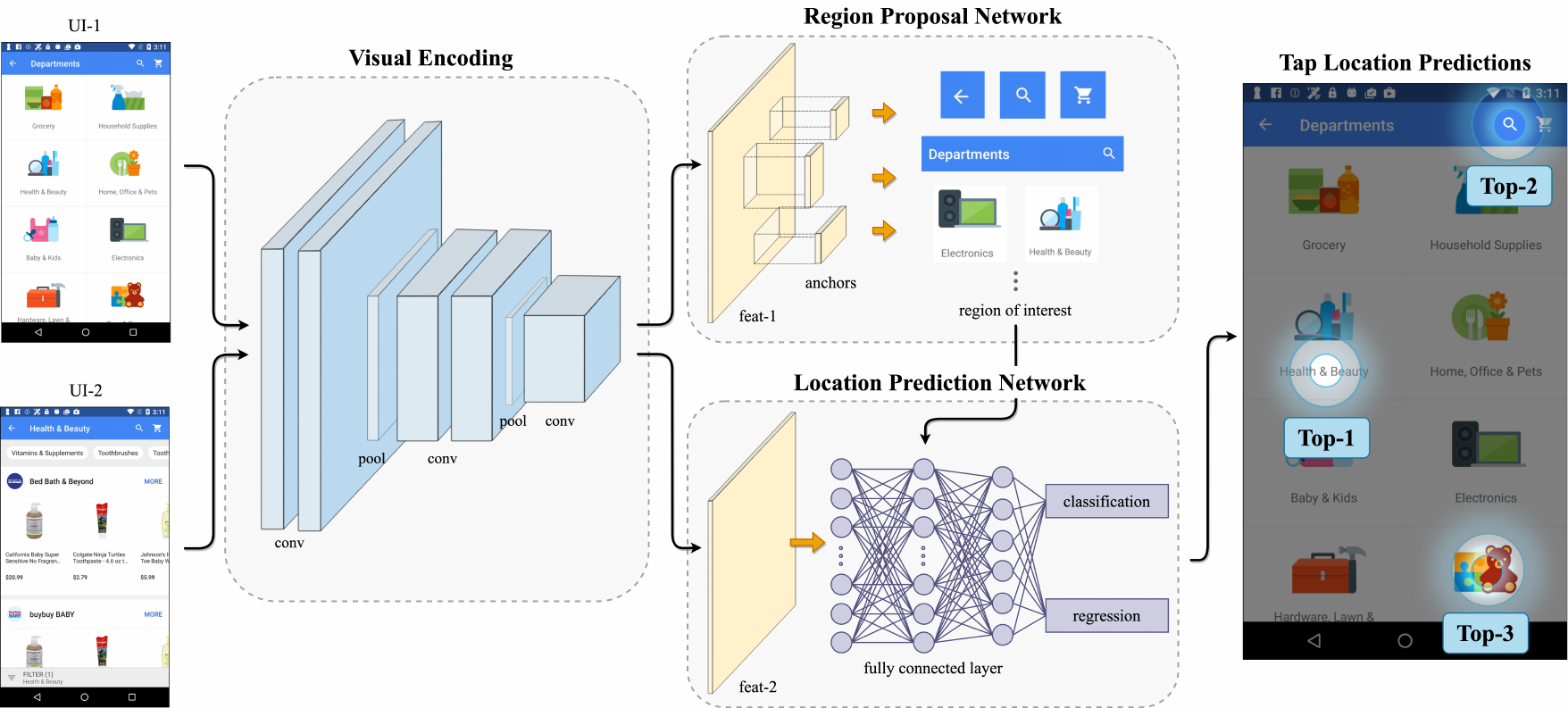}
	\caption{The model architecture to predict tapping locations.
	}
	\label{fig:tap_approach}
\end{figure*}

\subsection{Action Location Prediction}
\label{sec:location_prediction}
Different from \textsc{SCROLL} and \textsc{BACKWARD} actions, \textsc{TAP} action is sensitive to the action location as clicking the different buttons will trigger different functionalities of the app.
To accomplish this, we propose a novel deep-learning model that first recognizes the potential tappable area in the first UI and then predicts the tapping location that is perceived to transit to the second UI.
To increase the robustness of the deep-learning model, we propose UI-specific data augmentation methods to integrate human knowledge into the model, and a post-processing method to further improve the model predictions.

\subsubsection{Model Architecture}
Consider a UI transition (UI-1 $\xrightarrow[]{}$ UI-2), where UI-1 is the current UI, that transits to the next UI (UI-2).
The overview of our tapping location inference model is shown in Figure~\ref{fig:tap_approach}, which consists of three main components: \textit{Visual Encoding}, \textit{Region Proposal Network}, and \textit{Location Prediction Network}.

For \textit{Visual Encoding} of the feature map for images, we adopt the most commonly applied approach ResNet-101 (Residual Neural Network)~\cite{krizhevsky2012imagenet} with skip connections among the convolutional layers to capture more features in the image.
To accelerate the training process, we apply fine-tuning~\cite{tajbakhsh2016convolutional} on the pre-trained model ImageNet~\cite{krizhevsky2012imagenet} which already distills the visual features from more than 14 million natural images.
Specifically, we freeze the top few blocks of layers that store useful low-level features that can also apply to UI (e.g., edges, curves, etc.), but train the last block of layers to learn the higher-order UI feature representations (e.g., element, layout, etc.).

Inspired by the object detection task of detecting instances of objects of a certain class within a natural image, we exploit the neat network design \textit{Region Proposal Network (RPN)}~\cite{ren2015faster} to narrow down the feature maps by recognizing the perceived tappable areas in the UI.
In detail, given the feature map, RPN generates a set of region proposals (a.k.a anchor boxes) to compute the region of interest (RoI) scores to determine whether the regions contain tappable elements or not.
As the size and aspect ratio of elements in the UI is different from the objects in the natural scenes, we define five anchor-box scales (e.g., 32, 64, 128, 256 and 512), and four aspect width:height ratios (e.g., 1:1, 2:1, 4:1 and 8:1), which is empirically tested in UI element detection~\cite{chen2020object}.

Once we obtain the potential tappable areas in the UI-1, we propose \textit{Location Prediction Network} to predict the specific tap locations that transit to UI-2.
Considering the UI transition, we first jointly combine the feature map of the potential tappable areas detected in UI-1 and the feature map of UI-2.
Then, these features are given as input to a fully connected layer, whose output then goes into two branches. 
One branch for location regression is used to predict coordinate, and the other for classification that applies a Softmax activation layer to compute the probability of the coordinate to transit to UI-2. 
In the end, we output the inferred tapping coordinates in confident ranking order.

\subsubsection{Loss Function}
\label{sec:loss_function}
To train our proposed deep-learning model, we introduce a tailored loss function, which consists of \textit{classification loss} and \textit{regression loss}.
The \textit{classification loss} is to train the model when the probability of the predicted coordinate diverges from the ground-truth.
To achieve this, we leverage CrossEntropyLoss~\cite{murphy2012machine} to calculate the classification loss among 2 classes, where 0 indicates the predicted coordinate cannot trigger the UI transition, otherwise 1.
The \textit{regression loss} is to train the model when the predicted coordinate $(x,y)$ lies out of the ground-truth bound $(x_{lower},y_{lower},$ $x_{upper},y_{upper})$.
It is composed of the horizontal loss (x dimension) and vertical loss (y dimension) $Loss_{reg} = Loss_{reg_x} + Loss_{reg_y}$.
An example of regression loss from x dimension (likewise from y dimension) is calculated as \(Loss_{reg_x} = \mathds{1}_{\notin[x_{lower}, x_{upper}]}^{x} smooth_{L1}(x-\frac{x_{lower}+x_{upper}}{2})\),
where $\mathds{1}_{\notin[x_{lower}, x_{upper}]}^{x}$ is an indicator whose value is 1 if $x$ is out of the bound $(x_{lower},x_{upper})$.
$smooth_{L1}$ is the robust regression loss function Smooth L1~\cite{girshick2015fast}.
Usually, the boundary is loose and the key content is centered, therefore, we regress the coordinate towards the middle of the bound ($\frac{x_{lower}+x_{upper}}{2}$).

\begin{figure}
	\centering
	\includegraphics[width = 0.98\linewidth]{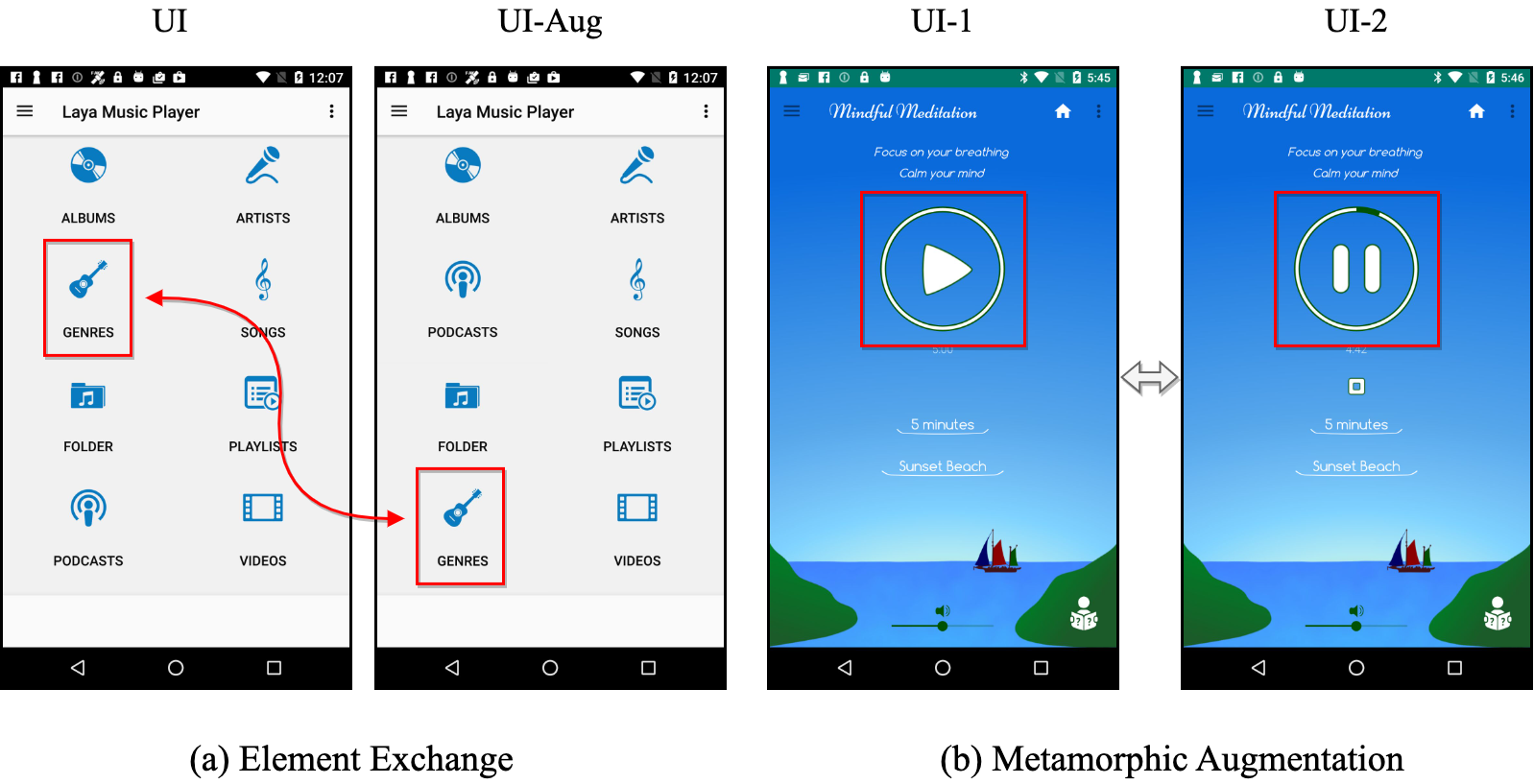}
	\caption{An illustration of data augmentation method.}
	\label{fig:augExample}
\end{figure}

\subsubsection{Data Augmentation}
\label{sec:augmentation}
The foundation of training deep-learning models is big data.
Although we label some actions in the app tutorial videos in our empirical study (Section~\ref{sec:background}), the set of actions is not sufficient and manual labeling is prohibitively expensive.
Therefore, we adopt one of the largest UI transition datasets Rico~\cite{deka2017rico}.
The Rico dataset contains 55k unique transition traces from 9.3k Android apps.
The transition trace is represented as a sequence of UI screens, as well as information about the interactive coordinate and element.
Rico also captures a video to record the transition trace.

While Rico has a large amount of UI transition data, it may not cover abundant tapping patterns discovered in our empirical study (Section~\ref{sec:pattern2}).
To integrate human knowledge into the model, we apply data augmentation which is a technique used to create new synthetic data from existing data based on heuristic patterns.
Specifically, we apply two UI-specific data augmentation methods, e.g., \textit{Element Exchange} and \textit{Metamorphic Augmentation}.

\textit{Element Exchange:} UI is not merely a collection of individual and unrelated elements, such as texts, images, buttons, etc.
Instead, it is designed with high-level semantics, forming perceptual groups such as tab, menu, card, or list.
To keep the UI design consistent, the elements in the perception group often look similar~\cite{xie2022psychologically}.
According to this observation, we apply \textit{Element Exchange} to generate a number of synthetic samples by switching the position of similar UI elements in the perception group, without affecting the nature of UI.
In detail, we first search the UIs in the Rico dataset that use certain Android layout classes that may contain a group of elements (e.g., ListView, FrameLayout, Card, TabLayout).
Then, we heuristically examine the elements in the group to filter out those are not similar by width, height, element class, etc.
For example, as shown in Figure~\ref{fig:augExample}(a), UI-Aug is artificially generated by switching the element ``GENRES'' to ``PODCASTS'' in the UI.

\textit{Metamorphic Augmentation:} Apart from augmenting the dataset based on a single UI, our task aims to predict the tapping location from one UI to another, prompting us to develop a tailored data augmentation method for pair of UIs.
Inspired by the metamorphic testing~\cite{chen2020metamorphic}, some of the UI transitions can be reversed by tapping on the same location.
For example, as shown in Figure~\ref{fig:augExample}(b), tapping on the ``play'' button in UI-1 will transit to the ``pause'' button in UI-2, and vice versa.
To achieve this, we search the UI transitions that tap on certain elements that have opposite semantics (e.g., ``play-pause'', ``on-off'' switch, ``selected-unselected'' checkbox, etc.), and therefore add reverse samples in the training dataset to help the model learn deep human knowledge.

\begin{figure}
	\centering
	\includegraphics[width = 0.945\linewidth]{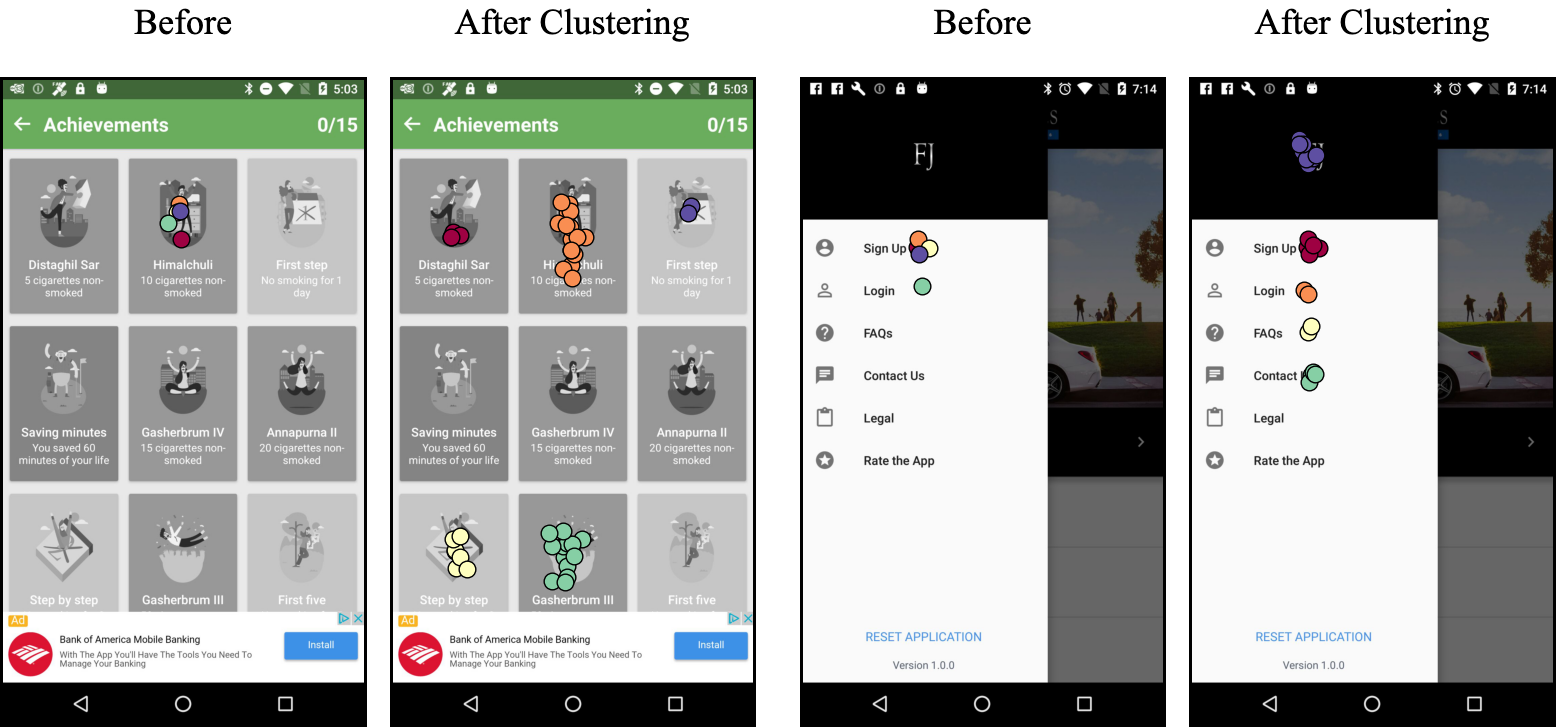}
	\caption{The top-5 candidate tapping positions with and without clustering.}
	\label{fig:cluster}
\end{figure}

\subsubsection{Post-processing}
\label{sec:post}
As there are many tapping coordinates predicted by the model, and some of them are very close to each other, which may affect the effect of action location recommendation as shown in Figure~\ref{fig:cluster}.
We further post-processing these inferred tapping coordinates by a clustering algorithm, density-based spatial clustering (DBSCAN)~\cite{schubert2017dbscan}, to drive more effective predictions.
In detail, DBSCAN finds the nearby coordinates by Euclidean distance to form clusters and iteratively expands if its neighbors are close.
Therefore, two parameters are required: the minimum number of points $min_{pts}$ and a point needs to have within a certain radius $\epsilon$ in order to be included in a cluster.
We set $min_{pts}$ to 1 and the value of $\epsilon$ to 40, empirically by a small-scale experiment.
Within each cluster, we choose the most confident coordinate as the representative of the cluster, yielding the tap location as shown in Figure~\ref{fig:cluster}.

\begin{figure*}
	\centering
	\includegraphics[width = 0.735\linewidth]{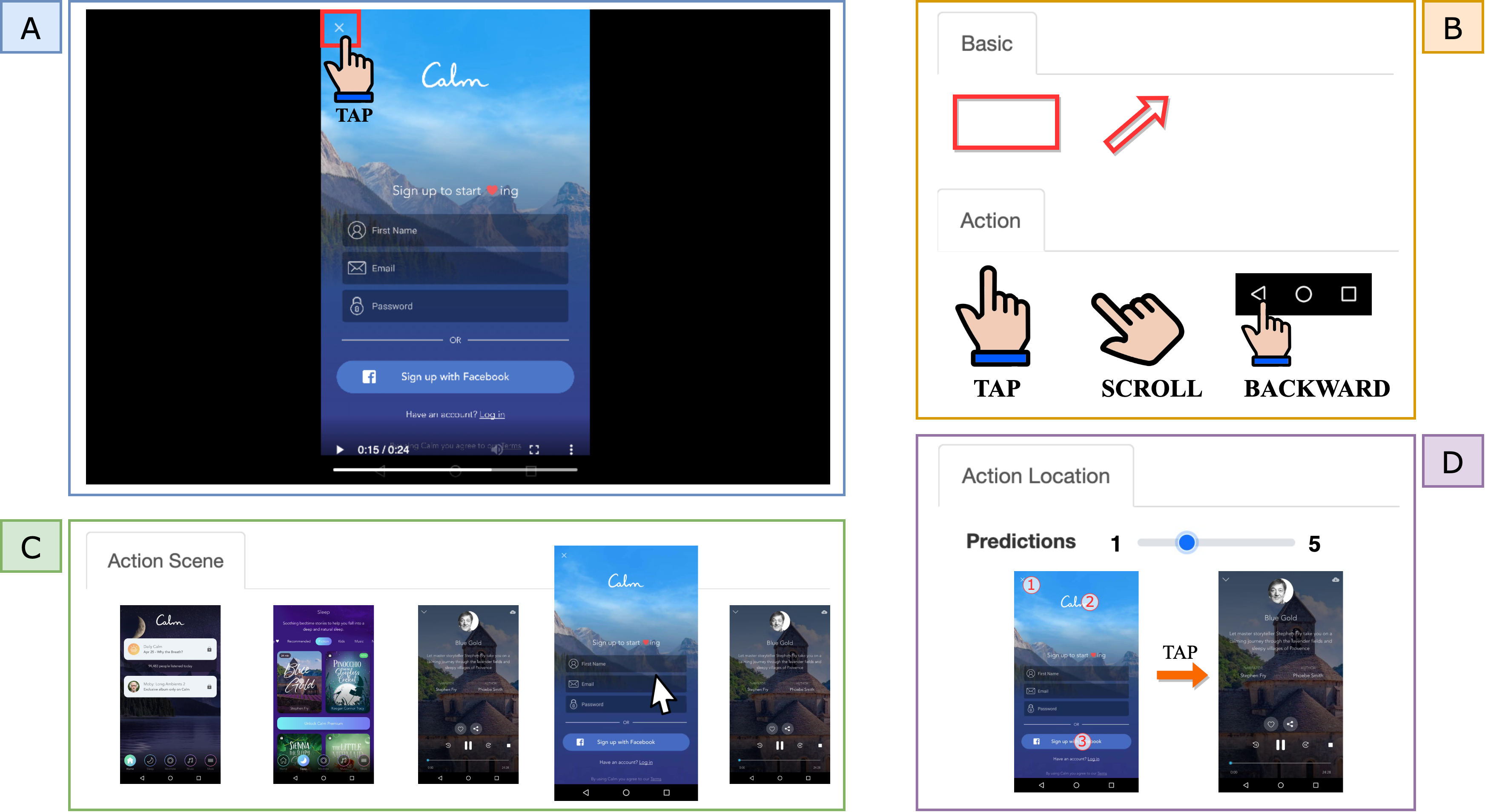}
	\caption{The implementation of \tool. The interface includes four major components: a video playback screen (A), an annotation box (B), an action scene box (C), and an action location box (D).}
	\label{fig:implementation}
\end{figure*}

\section{Implementation of \tool}
With the acquired action scenes and action locations, we build a proof-of-concept \tool to allow video creators to interactively access the generated action scenes, visualize the predicted action locations, and eventually create annotations.
The overview of the user interface is shown in Figure~\ref{fig:implementation}, including four interactive components.

\textbf{(1) Video Playback Screen}:
We allow the user to watch and annotate the app tutorial video in Figure~\ref{fig:implementation}A.
We also provide a playback slider for user to navigate to arbitrary frames in the videos.

\textbf{(2) Annotation Box}:
With the annotation box in Figure~\ref{fig:implementation}B, the user can interactively create and modify the annotation of an action.
We implement the annotation box by using Drag \& Drop API~\cite{web:drag}, so that the user can easily annotate the action by dragging the action kit and dropping it onto the video playback screen, as shown in Figure~\ref{fig:implementation}A.

\textbf{(3) Action Scene}:
As shown in Figure~\ref{fig:implementation}C, the action scenes are automatically generated by our approach in Section~\ref{sec:scene_generation}.
Each UI frame illustrates an action scene in the video.
By clicking on the UI frame, the video will jump directly to the timeline where the action takes place.

\textbf{(4) Action Location}:
To help the user efficiently identify the actions triggered to the next UI, we automatically predict the potential action locations in Figure~\ref{fig:implementation}D.
For \textsc{TAP} action, there are many locations predicted by our approach as discussed in Section~\ref{sec:location_prediction}.
We provide user with top-k predictions to calibrate the final action location.
Note we only consider k in the range 1-5, as the users rarely check a long recommendation list.

With the help of our \tool, the action annotation process is straightforward. 
The video creator first positions the video at a frame of interest (where action performs) by clicking the UI frame in the action scene box (Figure~\ref{fig:implementation}C).
According to the action locations in the recommendation list (Figure~\ref{fig:implementation}D), the video creator can examine frames back and forth in the video (Figure~\ref{fig:implementation}A) to quickly identify the real action location.
Finally, the video creator can leverage the action kits in the annotation box (Figure~\ref{fig:implementation}B) to annotate the action in the video frame.

\section{Automated Evaluation}
\label{sec:automated_eval}
In this section, we describe the procedure we use to evaluate each phase of our approach in terms of its performance automatically.

\subsection{Action Scene Generation}
\subsubsection{Testing Data}
To evaluate the ability of our approach to accurately segment the UI videos into action scenes, we utilize 6k UI videos from the Rico dataset~\cite{deka2017rico}.
Each video provides a sequence of actions as the ground-truth.
In total, we collect 30k \textsc{TAP} actions, 3k \textsc{SCROLL} actions, and 2k \textsc{BACKWARD} actions.
On average, a 30s UI video contains 6.29 actions.

\subsubsection{Baselines}
To demonstrate the advantage of using SSIM to segment scenes from the UI videos, we compare it with 4 widely-used image similarity metrics as baselines, including 2 pixel-level (e.g., \textbf{absolute differences}~\cite{watman2004fast}, \textbf{color histogram}~\cite{wang2010robust}) and 2 structural-level (e.g., \textbf{SIFT}~\cite{lowe2004distinctive}, \textbf{ORB}~\cite{rublee2011orb}).
In addition, we set up 3 state-of-the-art methods (e.g., image processing, and machine learning) which are commonly-used for video segmentation as the baselines to compare with our method.
\textbf{PySceneDetect}~\cite{web:pyscenedetect} is a practical Python library to detect shot boundaries by analyzing color, intensity, and motion estimation between frames.
\textbf{Hecate}~\cite{song2016click} is a tool developed by Yahoo to generate shot boundaries by estimating frame quality and using machine learning to cluster frames and aggregate them as shots.
\textbf{Scene Edit Detection}~\cite{web:sed} is a handy feature in Adobe Premiere Pro CC, that leverages machine learning to automatically detect cut points and scene changes from the video.

\subsubsection{Evaluation metrics}
We employ two widely-used evaluation metrics, e.g., Video F1-score, and Levenshtein score.
To evaluate the precision of detecting the shots from the UI videos, we adopt the Video F1-score~\cite{truong2021automatic}, which is a standard video shot boundary metric to measure the difference between two sequences of shots that properly accounts for the relative amount of overlap between corresponding shots.
Consider the shots detected by our approach ($c_{our}$) and ground-truth ($c_{gt}$), the Video F1-score is computed as $ \frac{2|c_{our} \cap c_{gt}|}{|c_{our}| + |c_{gt}|} $, where $|c|$ denotes the duration of the shot.
The higher the score value, the more precise the approach can detect the shots.
To evaluate the accuracy of generating action scenes, we adopt the Levenshtein score~\cite{navarro2001guided}, which compares the sequence of ground-truth actions and generated actions.
We express the score value in percentage.
The higher the score value, the more similar the generated action scene is to the ground-truth.
If the action scene generated by our approach exactly matches the ground-truth, the score value is 100\%.

\renewcommand{\arraystretch}{0.95}
\begin{table}
	\centering
        \small
        \tabcolsep=0.3cm
	\caption{Performance comparison of action scene generation.}
	\label{tab:scene_performance}
	\begin{tabular}{l|c|c} 
	    \hline
	    \bf{Methods} & \bf{Video F1-score} & \bf{Levenshtein} \\
	    \hline
	    Absolute~\cite{watman2004fast} & 63.41\% & 72.18\% \\
	    Histogram~\cite{wang2010robust} & 73.77\% & 76.34\% \\
	    SIFT~\cite{lowe2004distinctive} & 54.65\% & 63.33\% \\
	    ORB~\cite{rublee2011orb} & 53.92\% & 62.61\% \\
	    PySceneDetect~\cite{web:pyscenedetect} & 38.28\% & - \\
	    Hecate~\cite{song2016click} & 32.64\% & - \\
            Scene Edit Detection~\cite{web:sed} & 41.02\% & - \\
	    \bf{\tool} & \bf{81.67\%} & \bf{86.41\%} \\
		\hline
	\end{tabular}
\end{table}

\subsubsection{Results}
Table~\ref{tab:scene_performance} shows the overall performance of all baselines.
The performance of our approach \tool is much better than that of other baselines, i.e., 10.7\%, 13.1\% boost in Video F1-score, and Levenshtein score even compared with the best baseline (Histogram).
We observe that the state-of-the-art methods do not work well in our task, i.e., only achieve 38.28\%, 32.64\%, and 41.02\% in Video F1-score for PySceneDetect, Hecate, and Scene Edit Detection, respectively.
The issues with these methods are that they are designed for general videos which contain more natural scenes like humans, plants, animals, etc.
Different from those videos, the UI videos belong to artificial artifacts with different image motions (i.e., UI rendering).

We also observe that the pixel-level similarity methods (Absolute, Histogram) perform better than structural-level methods (SIFT, ORB), i.e., on average 14.3\% and 11.3\% improvement in Video F1-score and Levenshtein score, respectively.
This is because, unlike images of natural scenes, the keypoints/features in the UIs may not be distinct.
For example, a UI contains multiple identical checkboxes, and the duplicate keypoints of checkboxes can significantly affect similarity computation.

Although the method based on the pixel-level metric (Histogram) achieves the best performance in the baselines, it does not perform well compared to our approach, i.e., 73.77\% vs 81.67\% in Video F1-score, and 76.34\% vs 86.41\% in Levenshtein score.
This is because the color histogram is sensitive to the pixel value. 
The UI videos can often have image noise due to fluctuations of color or luminance, which may significantly affect pixel measurements.
In contrast, our approach using SSIM achieves better performance as it takes similarity measurements in many aspects from spatial and pixel, which allows for a more robust measurement.

Albeit the good performance of \tool, we still make wrong action scene generation for some UI videos due to two main reasons.
First, some UIs may contain animated app elements such as advertisements or movie playing, which will change dynamically, resulting in inaccurate shot detection.
Second, some UI videos start with the \textsc{BACKWARD} action, which limits our approach as we detect backward by comparing it with the previous UIs.

\subsection{Action Location Prediction}
\label{sec:eval_location}
Different from \textsc{SCROLL} and \textsc{BACKWARD} actions which are not sensitive to the action location, \textsc{TAP} location directly determines the response of the action.
Therefore, we systematically evaluate the performance of tapping location prediction in this section.

\subsubsection{Testing Data}
Since our approach employs a deep-learning model (Section~\ref{sec:location_prediction}) to predict the tapping location, we train and test our model using the Rico dataset as discussed in Section~\ref{sec:augmentation}.
Note that a simple random split cannot evaluate the model's generalizability, as tapping on the screens in the same app may have very similar visual appearances.
To avoid this data leakage problem~\cite{kaufman2012leakage}, we split the screens in the dataset by apps.
The resulting split has 26k (85\%) tapping actions in the training dataset, 2k (8\%) in the validation dataset, and 2k (7\%) in the testing dataset.
In addition, we apply the data augmentation methods in Section~\ref{sec:augmentation} to further enhance our training dataset.
In total, we create 26\% additional data, resulting in 33k samples in the training dataset.

\subsubsection{Baselines \& Ablations}
We set up 3 state-of-the-art methods which are widely-used for tap location prediction as the baselines to compare with our method.
\textbf{ActionBert}~\cite{he2021actionbert}, that first detects tappable elements in the UI, and then trains a classification model to identify which element is likely to trigger the tapping action. 
Since ActionBert is not publicly released, we follow their original paper to replicate the approach.
\textbf{Humanoid}~\cite{li2019humanoid} is another baseline, which proposes a recurrent neural network to predict how the user will tap the UI step-by-step.
Since our task predicts the tapping location based on a pair of UIs, we utilize the widely-used \textbf{Siamese} network~\cite{koch2015siamese} as our baseline, which encodes the visual information from pair of images to yield predictions.
Specifically, we use the state-of-the-art ResNet-101 architecture to capture the visual information (the same as our model) and predict two numeric variables corresponding to the tap location $(x,y)$.

To further demonstrate the advantage of our approach, we set up 3 ablation studies.
Since we propose a tailored loss function to optimize the model training in Section~\ref{sec:loss_function}, we consider a variant of our approach without the tailored loss function \textbf{\tool w/o loss} to see the impact of the loss optimization.
We further investigate the contribution of our data augmentation methods in Section~\ref{sec:augmentation}, namely \textbf{\tool w/o augmentation}, to see the performance of our model trained without 6,865 (26\%) additional data.
As discussed in Section~\ref{sec:post}, we propose a post-processing method to cluster the model predictions and filter the redundant ones.
Therefore, we consider a variant of \textbf{\tool w/o post-processing} to compare the performance of our approach with and without post-processing.

\subsubsection{Evaluation metrics}
We formulate the problem of tap location prediction as a searching task (i.e., search the most likely location to tap to the next UI), so we employ Precision@k to evaluate the performance.
As one UI element occupies a certain area, tapping any specific point within that area can successfully trigger the action.
Therefore, Precision@k is the proportion of the top-k predicted locations within the ground-truth UI element.
The higher value of the metric is, the better a search method performs.
Note we only consider k in the range 1-5, as users rarely check a long recommendation list.

\renewcommand{\arraystretch}{1}
\begin{table}
	\centering
        \small
	\caption{Performance comparison of \textsc{TAP} location prediction. ``Prec'' denotes the precision of the predicted tap locations.}
	\label{tab:tap_prediction_performance}
	\begin{tabular}{l|c|c|c} 
		\hline
		\bf{Methods} & \bf{Prec@1} & \bf{Prec@3} & \bf{Prec@5} \\
		\hline
		ActionBert~\cite{he2021actionbert} & 36.36\% & 41.60\% & 56.24\% \\
		Humanoid~\cite{li2019humanoid} & 29.72\% & 34.23\% & 47.22\% \\
		Siamese~\cite{koch2015siamese} & 22.32\% & - & - \\
		\hline
		\tool w/o loss & 46.58\% & 61.63\% & 71.33\% \\
		\tool w/o augmentation & 45.46\% & 63.92\% & 74.01\% \\
		\tool w/o post-processing & 46.21\% & 48.34\% & 51.77\% \\
		\bf{\tool} & \bf{50.14\%} & \bf{69.32\%} & \bf{81.89\%} \\
		\hline
	\end{tabular}
\end{table}

\subsubsection{Results}
Table~\ref{tab:tap_prediction_performance} shows the overall performance of all methods.
The performance of our model \tool is much better than that of other baselines in all metrics, i.e., 13.78\%, 27.72\%, 30.65\% higher in Precision@1, Precision@3, and Precision@5 even compared with the best baseline (ActionBert).
In contrast with the baselines, the Siamese network only achieves 22.32\% in Precision@1, which confirms the difficulty of predicting a specific tapping coordinate in the UI beyond a simple regression model. 
ActionBert adopts more advanced models to predict the tapping locations, but it still does not perform well compared to our model (36.36\% vs 50.14\% in Precision@1).
This is because, ActionBert applies a multi-phase pipeline that first detects the elements in the UI, then extracts their information, and finally predicts the tap locations.
This pipeline may lead to a ``garbage in and garbage out'' problem, i.e., imprecise UI element detection will result in incorrect tap location predictions.
In contrast, our model is end-to-end differentiable, which is more robust to predict the specific tap location.

We further demonstrate the advantage of our model with ablation studies in  Table~\ref{tab:tap_prediction_performance}.
We can see that applying the post-processing method can significantly improve the performance of our model, i.e., improving 3.93\%, 20.98\%, and 30.12\% in Precision@1, Precision@3, and Precision@5, respectively.
This is because, many of the predicted locations are very close to each other (as shown in Figure~\ref{fig:cluster}), resulting in redundant predictions.
Compared to traditional loss functions, our model with tailored loss optimization can achieve better performance, i.e., on average 7.27\% improvement in precision.
In addition, augmenting more training data improves 4.68\%, 5.4\%, and 7.88\% model performance in Precision@1, Precision@3, and Precision@5, respectively.
This suggests that our human knowledge-based data augmentation methods can further improve the model to capture the characteristics of UI transitions.

\begin{figure}
	\centering
	\includegraphics[width = 0.99\linewidth]{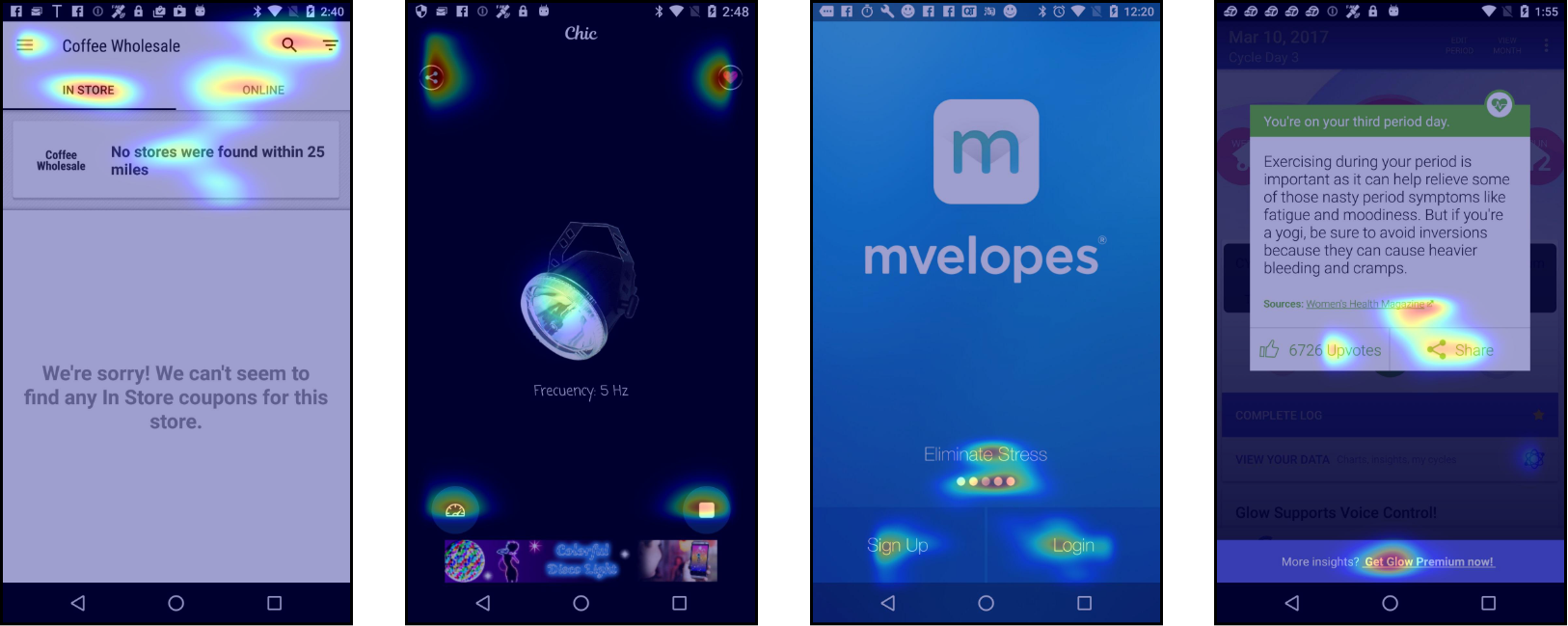}
	\caption{The heatmap of tappable locations in the UI predicted by our model.}
	\label{fig:heatmapExample}
\end{figure}

\begin{figure*}
	\centering
	\includegraphics[width = 0.90\linewidth]{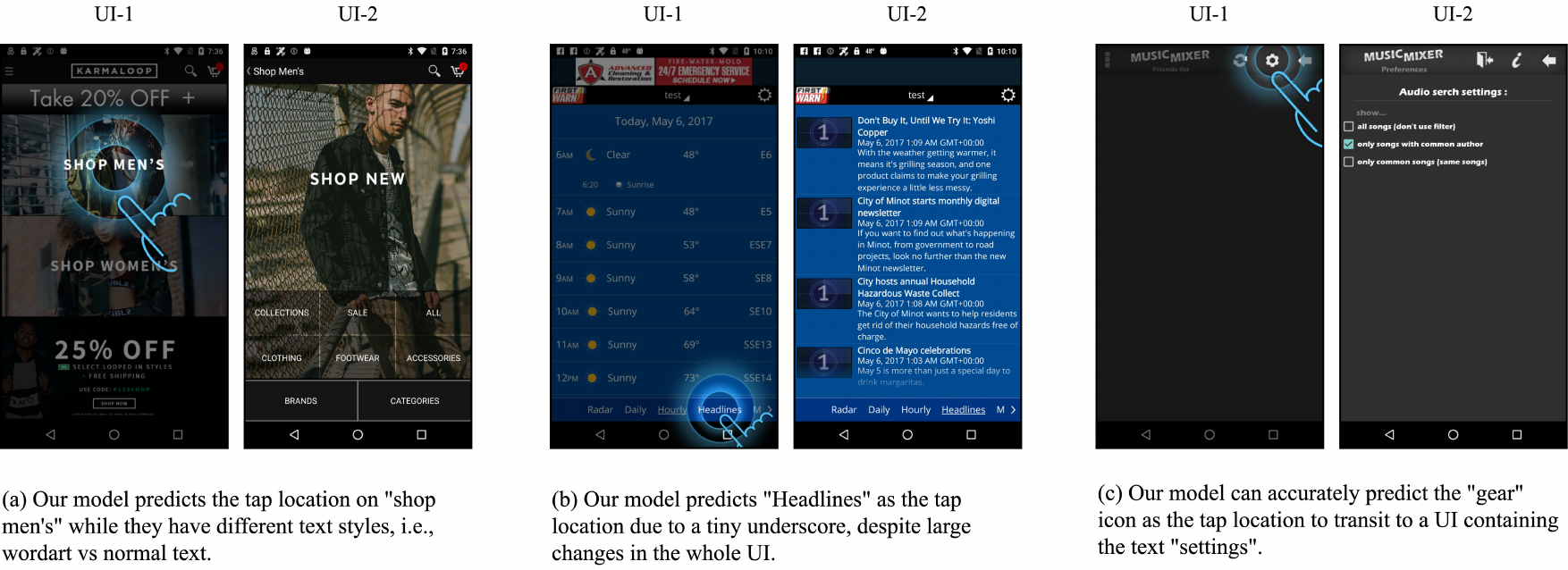}
	\caption{Examples of three kinds of accurately predicted tap locations. Blue color represents the prediction by our model.}
	\label{fig:rightExample}
\end{figure*}

To assess the trust of the model and interpret how the model gives a certain prediction, we visualize the features used to infer the final tap location into a heatmap by a visualization technique Grad-CAM~\cite{selvaraju2017grad}.
Figure~\ref{fig:heatmapExample} presents examples of the conclusive feature heatmap from our model.
We can see that most of the predicted tap locations are spotted on the tappable elements, indicating the reliability and interpretability of our model.
Figure~\ref{fig:rightExample} shows some predicted tap locations for UI transitions.
We can see that our model can accurately predict the locations in different complex transitions, including handling content in different styles, being sensitive to tiny features, and being robust to non-homology features.

\begin{figure*}
	\centering
	\includegraphics[width = 0.90\linewidth]{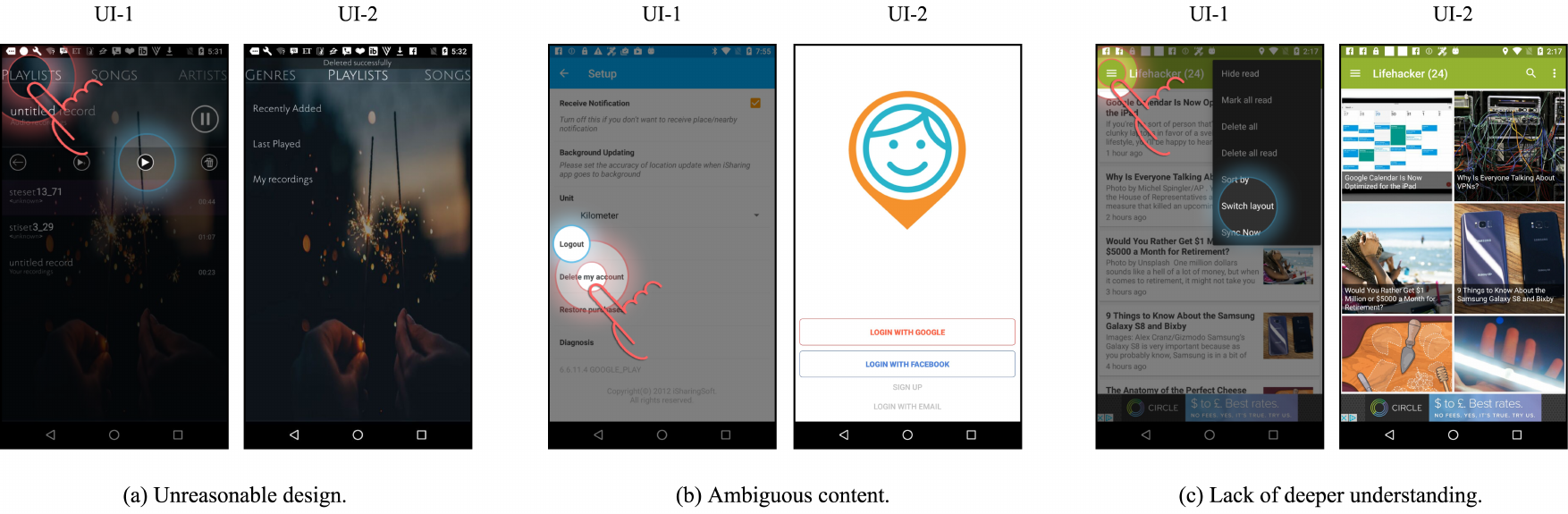}
	\caption{Examples of three prediction errors. Blue color represents the ground-truth, and red color represents the prediction by our model.}
	\label{fig:wrongExample}
\end{figure*}

For the failure analysis of our model, we conclude three main reasons why our model fails.
First, some UI designs are not reasonable in low-quality mobile apps.
For example, in Figure~\ref{fig:wrongExample}(a), the designer proposes a ``Play'' icon to transit to a ``Playlist'' UI, while our predicted ``Playlist'' button in the multi-tab is more aligned to user experience.
Second, the contents are ambiguous, such as ``Logout'' and ``Delete my account'' in Figure~\ref{fig:wrongExample}(b) may both be able to transit to the next UI.
Note that ``Logout'' is the second predicted tap location, suggesting this issue can be potentially solved by expanding the location search.
Third, some UI transitions require a deep semantic understanding of UIs.
For example, to predict the tap location in Figure~\ref{fig:wrongExample}(c), we need to first understand the contents in the UIs, and then through a difference analysis, we find that it is a structural change, so we speculate the semantic text ``Switch layout'' as the tap location.
In the future, we will improve the performance of our model by adding more semantic information, such as the contents, layout structures, etc.

\section{Usefulness Evaluation}
\label{sec:user_study}
We demonstrate the effectiveness of our approach in the last section, and we continue to show its usefulness with a user study to see if it can really assist video creators to annotate the actions in the app tutorial videos.

\subsection{Dataset of User Study}
We randomly select 8 app tutorial videos from Youtube, covering different app usage scenarios (i.e., financial, booking, systematic).
The details of the tutorial videos are shown in Table~\ref{tab:video_detail}, which consists of 4 short videos and 4 long videos.
On average, each short video is of 2.85 minutes and contains 15 actions; each long video is of 10.79 minutes and contains 49 actions.
Our approach achieves an average accuracy of 68.6\%, 82.9\%, and 93.1\% in predicting top 1, 3, and 5 actions, respectively.

\subsection{Experimental Design}
We recruit 12 video creators (8 females, 4 males) with experience in annotating app tutorial videos from an online posting. 5 from the app development team in the industry, 2 from the movie industry, and 5 freelancers who regularly post on video-sharing platforms. Their ages range from 24 to 36 years (M = 28.7, SD = 3.8).
Each participant will receive a \$50 shopping card as a reward after the experiment.
At the beginning of the study, we first give them an introduction to our study and also a demo tutorial video (not in the experimental dataset) to try.
We also conduct a follow-up survey among the participants regarding their annotation experience.

\renewcommand{\arraystretch}{1.05}
\begin{table*}[htp!]
    \small
	\centering
	\caption{The details of our experimental app tutorial videos.}
	\label{tab:video_detail}
	\tabcolsep=0.25cm
	\begin{tabular}{l|lcc|ccc} 
	\hline
	\bf{Video Group} & \bf{Title} & \bf{Length(min)} & \bf{\# TAP/SCROLL/BACK} & \bf{Prec@1} & \bf{Prec@3} & \bf{Prec@5}\\
	\hline
	\multirow{4}{*}{Short Video} & How to check screen time on Android? & 1.55 & 6 / 2 / 0 & 62.5\% & 87.5\% & 100\% \\ 
	& How to use Fitness \& Bodybuilding app? & 2.35 & 10 / 3 / 3 & 75\% & 87.5\% & 93.8\% \\
	& How to fix stopped Android apps? & 2.81 & 14 / 2 / 4 & 75\% & 90\% & 95\% \\
	& How to track usage in Edge app? & 4.68 & 13 / 4 / 1 & 61.1\% & 72.2\% & 83.3\% \\
	\hline
	\multirow{4}{*}{Long Video} & How to book an Airbnb? & 8.50 & 14 / 15 / 4 & 69.7\% & 78.8\% & 90.9\% \\
	& How to save Android battery life? & 9.62 & 34 / 14 / 11 & 61.1\% & 74.6\% & 86.4\% \\
	& How to manage budgets in Mint app? & 10.28 & 21 / 13 / 6 & 75\% & 92.5\% & 100\% \\
	& How to become a driver in Doordash app? & 14.78 & 47 / 4 / 14 & 69.2\% & 80\% & 95.4\% \\
	\hline
	\end{tabular}
\end{table*}

Participants are then asked to annotate 8 app tutorial videos in the experimental dataset individually in a quiet room, such as a lab or home, to minimize distractions.
The study involves two groups of six participants: the experimental group from $P_1$ to $P_6$ who gets help with the actions inferred by our approach, and the control group $P_7$ to $P_{12}$ who starts from scratch.
Each pair of participants $\langle P_x$ , $P_{x+6}\rangle$ has comparable annotation experience, so the experimental group has similar capability to the control group in total.
Our approach can produce some action prediction errors, but we do not carefully correct these errors or tell the participants which predictions are incorrect.
This is done to investigate the practical usefulness of our approach.
Participants are asked to finish each annotation as fast as they can while ensuring annotation accuracy.
To reduce stress bias, we allow them to take short breaks between each tutorial.
We record the time used to annotate the tutorial videos.
At the end of the session, we provide a cumulative questionnaire with 5-point Likert scale questions and a 5-minute open-ended interview to collect their feedback, in terms of the ease of annotations and the helpfulness of \tool.

\renewcommand{\arraystretch}{1.2}
\begin{table}
        \small
	\centering
        \tabcolsep=0.1cm
	\caption{Results for the questionnaires (Median, Inter-quartile Range).}
	\label{tab:questionnaire}
	\begin{tabular}{p{0.525\linewidth}p{0.3\linewidth}p{0.1\linewidth}} 
	    Statement & \multicolumn{2}{r}{Median (IQR)} \\
		\hline
		\rule{-8pt}{4ex}
		\bf{1.Easy to annotate} & & \\
		\rule{-8pt}{0ex}
		\footnotesize{1.1 I enjoyed the experience.} & \includegraphics[width=1\linewidth, height=0.16cm]{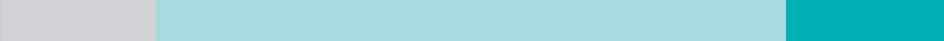} & \footnotesize{4.0(0.25)} \\
		\rule{-8pt}{0ex}
		\footnotesize{1.2 I was able to focus on the video.} & \includegraphics[width=1\linewidth, height=0.16cm]{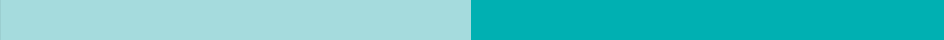} & \footnotesize{4.5(1.00)} \\
		\rule{-8pt}{0ex}
		\footnotesize{1.3 The mental effort required to annotate the actions was low.} & \includegraphics[width=1\linewidth, height=0.16cm]{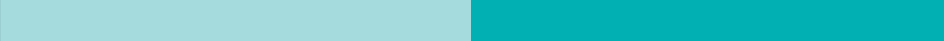} & \footnotesize{4.5(1.00)} \\
		
		\rule{-8pt}{3ex}
		\bf{2.Helpfulness} & & \\
		\rule{-8pt}{2ex}
		\footnotesize{2.1 It was helpful to reveal the action scenes.} & \includegraphics[width=1\linewidth, height=0.16cm]{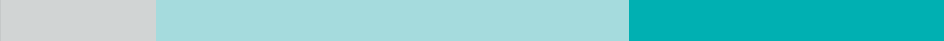} & \footnotesize{4.0(0.75)} \\
		\rule{-8pt}{0ex}
		\footnotesize{2.2 It was helpful to reveal what kinds of actions.} & \includegraphics[width=1\linewidth, height=0.16cm]{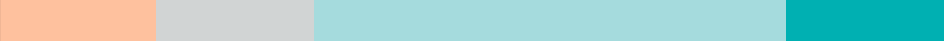} & \footnotesize{4.0(0.75)} \\
		\rule{-8pt}{0ex}
		\footnotesize{2.3 It was helpful to reveal where to touch.} & \includegraphics[width=1\linewidth, height=0.16cm]{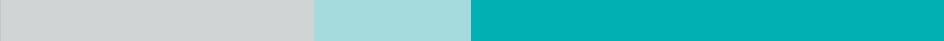} & \footnotesize{4.5(1.75)} \\
		\hline
		\multicolumn{3}{l}{
		    \begin{tabular}{p{0.05\linewidth}ccccccc}
             & \footnotesize{Strongly Disagree} &  \includegraphics[height=0.15cm, width=0.225cm]{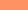} \footnotesize{1} &  \includegraphics[height=0.15cm, width=0.225cm]{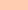} \footnotesize{2} &  \includegraphics[height=0.15cm, width=0.225cm]{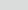} \footnotesize{3} &  \includegraphics[height=0.15cm, width=0.225cm]{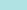} \footnotesize{4} &  \includegraphics[height=0.15cm, width=0.225cm]{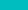} \footnotesize{5}
             & \footnotesize{Strongly Agree}
             \end{tabular}
		}
		
	\end{tabular}
\end{table}

\subsection{Results}
Overall, participants appreciate the usefulness of our \tool for revealing the actions performed in the app tutorial videos, so that they can easily annotate them. 
We present the annotation time and questionnaire results in Figure~\ref{fig:userstudy}.
The detailed questionnaire results for the experimental group are in Table~\ref{tab:questionnaire}.
To further understand the significance of the differences, we carry out the Mann-Whitney U test~\cite{fay2010wilcoxon} on the annotation time and questionnaire results between the experimental and the control group respectively. The test results suggest that our approach does significantly outperform the baseline in terms of these metrics with $p < 0.01$.

\subsubsection{Participant Behaviors}
\begin{figure*}[t!]
	\centering
	\includegraphics[width = 0.83\linewidth]{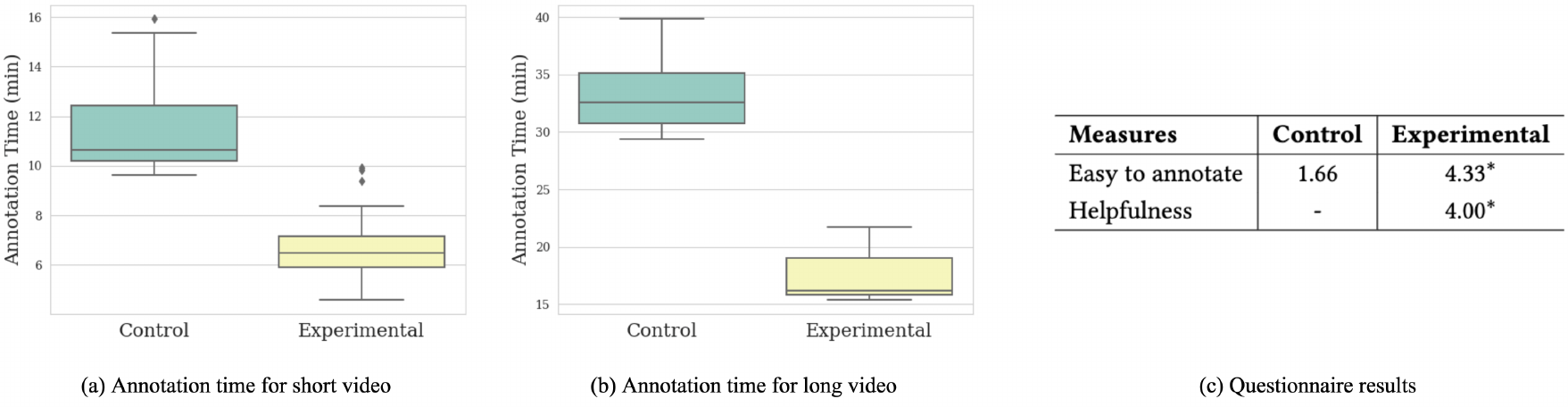}
	\caption{Performance comparison between the control and experimental group. $^*$ denotes \textit{p} $<$ 0.01.}
	\label{fig:userstudy}
\end{figure*}

As shown in Figure~\ref{fig:userstudy}(a) and Figure~\ref{fig:userstudy}(b), participants in the experimental group can annotate the actions much faster than the control group (with an average of 12.11 minutes vs 22.39 minutes).
That is the biggest strength of our approach, helping video creators annotate the actions in the app tutorial videos efficiently.
Specifically, with the help of our approach, 72\% and 90\% of the time are saved for annotating short videos and long videos, respectively.
This indicates the time savings become more evident for more actions and longer videos.

We also analyze the event logs from the study to gain a better understanding of participants' annotation processes.
We pay special attention to the falsely generated actions and find that none of these false actions were annotated by the participants, suggesting that participants can easily discern the correctness of predictions.

\subsubsection{Easy to Annotate}
Overall, participants respond that our tool \tool is easier to annotate the videos, e.g., 4.33 vs 1.66 compared to the annotation from scratch.
The questionnaire in Figure~\ref{tab:questionnaire} shows that participants in the experimental group enjoyed the experience (Q1.1). 
Five (83\%) of the participants agree that our interface is easy to understand and the annotation process is straightforward.
All of the participants in the experimental group are able to focus on the video (Q1.2).
One participant in the experimental group (P3) explains that 
\textit{``Typically, I have to check the video frame by frame to find a specific event. Longer videos require more effort, which can easily lead to distraction. With the help of the interactive action scenes in the tool, I can navigate directly to each event, saving me a lot of effort.''}
Participants in the experimental group also report lower mental effort (Q1.3) when annotating the actions in the videos, while three (50\%) of the participants in the control group express difficulty.
P7 in the control group says \textit{``The videos are too fast. It is really difficult to recognize the action in a short time. As a result, I have to replay the action many times, and carefully identify where the action performs.''}

\subsubsection{Helpfulness}
Participants find our tool helpful (4.00) in navigating actions (Q2.1), informing action types (Q2.2), and indicating action locations (Q2.3).
Five (83\%) of the participants in the experimental group praise the helpfulness of our generated action scenes (Q2.1).
On the one hand, it helps video creators to navigate directly to the frames when performing actions, improving annotation efficiency.
On the other hand, it can be helpful to understand the key actions in the video.
P4 and P6 mention that \textit{``The generated action scenes are particularly useful to me. This allows me to know in advance how many actions need to be annotated, and roughly the action flow of the video.''}
P3 further supports the usefulness of action scene generation in practice, \textit{``Youtube recommends the video creators add timestamps to their videos, representing the key moments. For the app tutorial videos, the key moments are just the action scenes, which can be automatically generated by \tool.''}
Most participants find that the action locations predicted by our tool contribute to their positive experience (Q2.3). 
P2 mentions that \textit{''Sometimes, the action locations are not inconspicuous to realize, especially those that don't have any animation effects like ripple, expand, etc.
It leaves me guessing the action locations.
In contrast, the action locations predicted by the approach can potentially provide hints for locating regions of interest.''}

\section{Discussion}
\tool has several opportunities for improvement.
First, while our approach saves 81\% of the time for annotating app tutorial videos, it still requires some manual effort from users, as our approach cannot achieve 100\% accuracy in inferring actions as discussed at the end of each subsection of the evaluation in Section~\ref{sec:automated_eval}. 
In the future, we aim to further improve the performance of our approach to minimize human interaction in the video annotation process as much as possible.
Second, we focus only on the most fundamental and common actions found in app tutorial videos. 
There are numerous other actions, such as pinch and rotate, which we believe can be addressed with a reasonable engineering effort. 
For some high-level gesture actions in 3D and AR/VR apps, a systematic study of patterns may be necessary.
Besides, we discuss the generality and the implications of our approach and put them as future work.

\subsection{Generalization of \tool}
\tool is designed to assist the action annotation of app tutorial videos to reduce the cognitive and interaction burdens of video creators in the annotation.
It has achieved satisfactory performance in generating action scenes and predicting action locations from Android app videos as evaluated in Section~\ref{sec:automated_eval}.
In addition to Android, there are also many other platforms such as iOS, web.
Supporting the videos of different platforms can bring analogous benefits to video creators.
For mobile platforms like iOS, the actions and usage patterns exert almost no difference from Android.
Therefore, our approach might be easily adapted to it with reasonable engineering effort.
For platforms using different devices like desktop, the differences between these platforms with Android can be considerably big.
In such cases, a detailed empirical study of the user behaviors and customization of our approach is required to determine the extension.
In the future, we will try to extend our approach to help video creators annotate the app tutorial videos of multi-platform.

\subsection{Bug Replay}
Bug recordings of mobile applications are easy to capture and include a wealth of information (i.e., reproduce steps, configurations, etc.), making them a popular mechanism for users to inform developers of the problems encountered in the bug reports.
In order to effectively resolve the bugs, the developer has to first understand the action steps performed in the bug recordings and then manually repeat them in the order shown.
This process can be time-consuming and error-prone, especially for novice developers~\cite{feng2022gifdroid}.
We would expect that our approach can be applied to bug recordings to extract the bug reproduction steps (i.e., a sequence of actions).
Once we derive the steps, we could further proceed to generate the testing script using Sikuli ~\cite{yeh2009sikuli} to automate bug replay.

\subsection{Video Captioning}
Captions or subtitles are provided to add clarity of details, better engage users, and translate the different languages~\cite{gernsbacher2015video}.
It is particularly useful for people with vision impairments (e.g., the aged or blind) to access the video content without requiring caregivers~\cite{liu2021makes}.
Our approach could be applied to enhance the accessibility of the app tutorial videos by generating clear and concise captions for action steps, enabling people with vision impairments to easily access the information and service of the mobile apps for convenience.

To that end, given UI scenes and action locations generated by our \tool, we could leverage the existing mature methods~\cite{zhang2021screen} to recognize the interacted UI element and detect any associated text.
Then, we could convert these UI elements into easy-to-understand natural language descriptions and embed them as subtitles.
The combination of video and text should provide a well-rounded and comprehensive learning experience.

\section{Conclusion \& Future Work}
This paper proposes \tool, a lightweight approach to support action annotation for app tutorial videos.
This approach uses image-processing and deep-learning methods to automatically generate the action scenes and predict the action locations.
We set up automated evaluations to demonstrate the performance of our approach, significantly outperforming the commonly-used and state-of-the-art baselines.
We further conduct a user study on the proof-of-concept interface to demonstrate the usefulness of \tool in helping video creators locate, analyze, and annotate actions more efficiently.

In the future, we will work in three directions.
First, we will keep improving our approach for better performance, such as incorporating the information of animation between UI transitions.
Second, we will develop our approach to support more high-level actions, such as pinch-in, gesture, etc.
Third, according to the user feedback, we will integrate our approach into the existing video editing tools, strengthening the collaboration between human and machine computation powers.

\bibliographystyle{ACM-Reference-Format}
\bibliography{main}

\end{document}